\documentclass[preprint, 3p, twocolumn]{elsarticle}
\usepackage{graphicx}
\usepackage{booktabs}
\usepackage{tabularx}
\usepackage[normalem]{ulem}
 \useunder{\uline}{\ul}{}
 \usepackage{rotating}
 \usepackage{dsfont}
 \usepackage{textcomp}
 \usepackage{upgreek}

\usepackage{color}
\def\Manu#1{\textcolor{black}{#1}}

\usepackage[switch]{lineno}
\usepackage{graphicx,xcolor}
\usepackage{epstopdf}
\usepackage{setspace}
\usepackage[T1]{fontenc}
\usepackage{amssymb,amsfonts,amsmath}
\usepackage{setspace}
\usepackage{subcaption}
\usepackage{mathtools}
\usepackage{etoolbox}
\biboptions{sort&compress}

\newcommand{\bit}[1]{\textit{\textbf{#1}}}

\usepackage{titletoc}
\titlecontents{section}
[0pt]                                               
{}%
{\contentsmargin{0pt}                               
    \thecontentslabel\enspace%
    \normalsize}
{\contentsmargin{0pt}\large}                        
{\titlerule*[.5pc]{.}\contentspage}                 

\patchcmd{\maketitle}
  {\finalMaketitle}
  {\finalMaketitle\tableofcontents\vspace{\baselineskip}}
  {}{}

\usepackage{hyperref}
\hypersetup{
    colorlinks=true,
    linkcolor=blue,
    filecolor=magenta,      
    urlcolor=cyan,
}
\begin{document}

\title{\Manu{Single bubble and drop techniques for characterizing foams and emulsions}}
\author[vinny]{V. Chandran Suja}
\ead{vinny@stanford.com}
\author[vinny,mariana]{M. Rodríguez-Hakim}
\author[vinny,javi]{J. Tajuelo}
\author[vinny]{G.G. Fuller}\ead{ggf@stanford.com}
\address[vinny]{Department of Chemical Engineering, Stanford University, Stanford, California 94305, USA}
\address[mariana]{Department of Materials, ETH Zurich, Vladimir-Prelog-Weg 5, Zurich 8093, Switzerland}
\address[javi]{Departamento de Física Interdisciplinar, Universidad Nacional de Eduación a Distancia UNED, Madrid 28040, Spain}

\begin{abstract}
The physics of foams and emulsions has traditionally been studied using bulk foam/emulsion tests and single film platforms such as the Scheludko cell. Recently there has been a renewed interest in a third class of techniques that we term as single bubble/drop tests, which employ isolated whole bubbles and drops to probe the characteristics of foams and emulsions. Single bubble and drop techniques provide a convenient framework for investigating a number of important characteristics of foams and emulsions, including the rheology, stabilization mechanisms, and rupture dynamics. In this review we provide a comprehensive discussion of the various single bubble/drop platforms and the associated experimental measurement protocols including the construction of coalescence time distributions, visualization of the thin film profiles and characterization of the interfacial rheological properties. Subsequently, we summarize the recent developments in foam and emulsion science with a focus on the results obtained through single bubble/drop techniques. We conclude the review by presenting important venues for future research.
\end{abstract}


\maketitle

\section{Introduction}\label{sec:Introduction}
Foams and emulsions are dispersions of a gas and a liquid, respectively, in a different liquid. Foams are common and desirable in a number of applications such as food manufacturing processes, personal and health care product development, detergency, firefighting, flotation of minerals, and waste water treatment \cite{frostad2016coalescence,pugh2016bubble, garrett2016science,patino2008implications,qian2009study,evans2002don,kitchener1984froth}. In contrast to these applications, stable foams are undesirable and need to be controlled in situations such as lubrication, textile dyeing, fermentation, and pulp and paper production \cite{suja2018evaporation,suja2020foam,sawicki1992high, tsang2006novel}. Emulsions are equally common and find important applications in food manufacturing processes, personal and health care product development, enhanced oil recovery, paints, pharmacy, and road construction \cite{leal2007emulsion,chappat1994some,mandal2010characterization,tadros2011colloids}. In contrast to these applications, stable emulsions are undesirable and have to be controlled during lubrication, desalting of crude oil, and fractionation of petroleum products \cite{harika2011impact, bochner2017droplet, azim2011demulsifier, umar2018review}. 

Motivated by the need to control the stability of foams and emulsions for different applications, researchers have developed a wide variety of experimental platforms to study the stability of these colloidal systems. Broadly, the existing experimental platforms can be classified as bulk foam/emulsion setups, single film setups and single bubble/drop setups. 

Bulk foam and emulsion experiments probe the physics of these colloidal systems at a bulk scale. They best mimic real life foams and emulsions, and capture all their complexity including many body interactions, the effects of advection, and the presence of plateau borders. Other advantages include the ease of operation and the convenient measurement of the aggregate properties. Common bulk foam tests include the industry standard ASTM D892 \cite{astm2013standard}, the foam rise test \cite{denkov2004mechanisms}, the shake test \cite{denkov2004mechanisms}, and the Flender foam test \cite{flenderfoamISO}. Common bulk emulsion tests include the industry standard ASTM D1401 \cite{astm2019standard}, shake test \cite{kamkar2020polymeric}, high pressure homogenization (microfluidization)  \cite{lobo2002coalescence,leal2007emulsion}, and membrane emulsification tests \cite{joscelyne2000membrane}. Detailed reviews on bulk foam and emulsion experiments and the corresponding characterization techniques are available in the literature \cite{ pugh1996foaming, friberg2010foams, pugh2005experimental, denkov2004mechanisms, fameau2017non, exerowa1997foam, mcclements2007critical, joscelyne2000membrane}. Despite their advantages, these tests are not suitable to systematically probe stabilization mechanisms due to the shear complexity of bulk foams and emulsions. Simplifications like 2D foams do exist \cite{miralles2014foam}, however, these systems are still inconvenient for developing a detailed understanding of the stability of thin liquid films that ultimately sustain foams and emulsions.  To overcome the limitations of bulk tests, researchers have developed a couple of other techniques. 

Single film techniques - the simplest abstraction of foams and emulsions - probe the stability of films that are analogous to those formed when two particles of the dispersed phase (gas bubbles or liquid drops) come close to each other \cite{exerowa2018foam}. Perhaps the most well known single film setup is the Scheludko-Exerowa cell, which was originally developed by Derjaguin and subsequently improved by Scheludko and Exerowa \cite{sheludko1967thin,exerowa1997foam}. Other variants include the Exerowa-Scheludko
porous plate cell and the Dippanear cell \cite{exerowa2008emulsion, dippenaar1982destabilization, garrett2016science}. Single film experiments have transformed our understanding of thin film stability. In particular, due to the ability of the technique to measure the pressure in the film, usually through a Scheludko-Exerowa cell, a deep understanding of the role of disjoining pressure in terminal thin film drainage and thin film stability has been developed \cite{stubenrauch2003disjoining}. Further, single film results have also aided in improving the theoretical understanding of thin film drainage, as the inherent reflection symmetry \cite{exerowa2018foam} in these experiments have made them attractive for theoretical and numerical analyses  \cite{joye1994asymmetric}. Detailed reviews on single film setups and results are available in the literature \cite{exerowa1997foam, langevin2015bubble}. Despite the above advantages, single film experiments have certain limitations including difficulties in conveniently controlling the size of the film and the approach velocity of the interacting interfaces, and the inability to study the interaction of interfaces with different radii of curvature.

To address these limitations and complement single film experiments, researchers have developed a third class of experimental tools that, in terms of mimicking real life foams and emulsions, fall midway between bulk tests and single film tests. These are referred to as single bubble/drop setups and, as their name indicates, utilize complete bubbles and drops to respectively understand foam and emulsion stability \cite{chan2011film}. Single bubble/drop experiments have three notable advantages over single film tests.  Firstly, single bubble/drop experiments allow the use of a complete bubble/drop, thus enabling the effects of the dispersed phase size and the rise velocity \cite{frostad2016coalescence} to be independently studied.  Secondly, single bubble/drop experiments can probe the interaction of interfaces with different radii of curvature, and have notably improved the understanding of coalescence at flat liquid-air interfaces \cite{suja2018evaporation,suja2020foam,kannan2018monoclonal}. Thirdly, in situ interfacial rheology measurements can be conveniently performed in single bubble/drop setups, thus making them a more holistic tool for developing a  mechanistic understanding of thin film stability \cite{kannan2018monoclonal,kannan2019linking}.

In this manuscript, we provide a comprehensive review of this important technique along with the recent developments in foam and emulsion science that came about through single bubble/drop experiments. We start with a brief discussion of the history of single bubble/drop experiments in \bit{Section \ref{sec:history}}. Subsequently, in \bit{Section \ref{sec:Methods}} we describe the single bubble and single drop setups in detail along with the relevant experimental protocols and data analysis techniques. In \bit{Section \ref{sec:FoamStability}} and \bit{Section \ref{sec:EmulsionStability}} we present the recent developments in foam and emulsion science, respectively. Finally, we conclude the manuscript by presenting important venues for future research.

\section{Historical Perspective}\label{sec:history}
In this section we will provide a brief overview of the important historical developments in the field of single bubble and single drop experiments. Comprehensive historical details on single film can be found in Gochev et al. \cite{gochev2016chronicles}. 

The early scientific interest in soap bubbles can be traced back to experiments performed by Boyle and Hooke \cite{plateau1873statique}. Initial scientific attention was focused on understanding the origin of the colors on soap bubbles. Notably, Newton performed experiments showing that the first bright color corresponds to a thickness of $107\;nm$ \cite{gochev2016chronicles} - a remarkably accurate result (see Fig.\ref{fig:Interferograms}). Subsequently, pioneered by the efforts of Plateau, the attention shifted to understanding the shape, interfacial properties, and stability of soap films. 

Investigations into the shape of soap films had a profound impact in the fields of differential geometry, calculus, and mechanics.  Notably, the field of \textit{Calculus of Variations} came about in part due to efforts by Bernoulli and his student Euler in the early 1700's to understand minimal surfaces formed by soap films \cite{isenberg1981soap}. The research into the shape of soap bubbles also resulted in the development of the famed Young-Laplace equation, the consequences of which were demonstrated elegantly by Charles Vernon Boys in a number experiments to the public \cite{boys1890soap}.  These advancements also paved the way for the development of the Axisymmetric Drop Shape Analysis (ADSA) and the Maximum Bubble Pressure Method (MBPM), two of the common techniques used to measure the surface tension using a bubble/drop supported on a capillary. ADSA was developed as a result of efforts since the late 1800's, notably by Worthington \cite{worthington1881ii,edgerton1937studies}, to utilize the shape of pendant liquid drops as the means to measure the interfacial tension. Over the years, as a result of advances in imaging and in computational methods, ADSA has become an indispensable tool for measuring interfacial tension \cite{berry2015measurement}. The first documented work on MBPM was reported by Simon in 1951 \cite{simon1851recherches}. Over the years as result of the efforts of number of researchers, MPBM is one of the most popular techniques to measure dynamic surface tension \cite{mysels1990maximum}. Interestingly, the famed quantum mechanist Erwin Schrodinger in 1915 provided the first accurate correction for MBPM measurements where the effects of gravity cannot be neglected \cite{schrodinger1915notiz}, before developing the other equation he is now known for.

\begin{figure*}[!th]
\includegraphics[width=\linewidth]{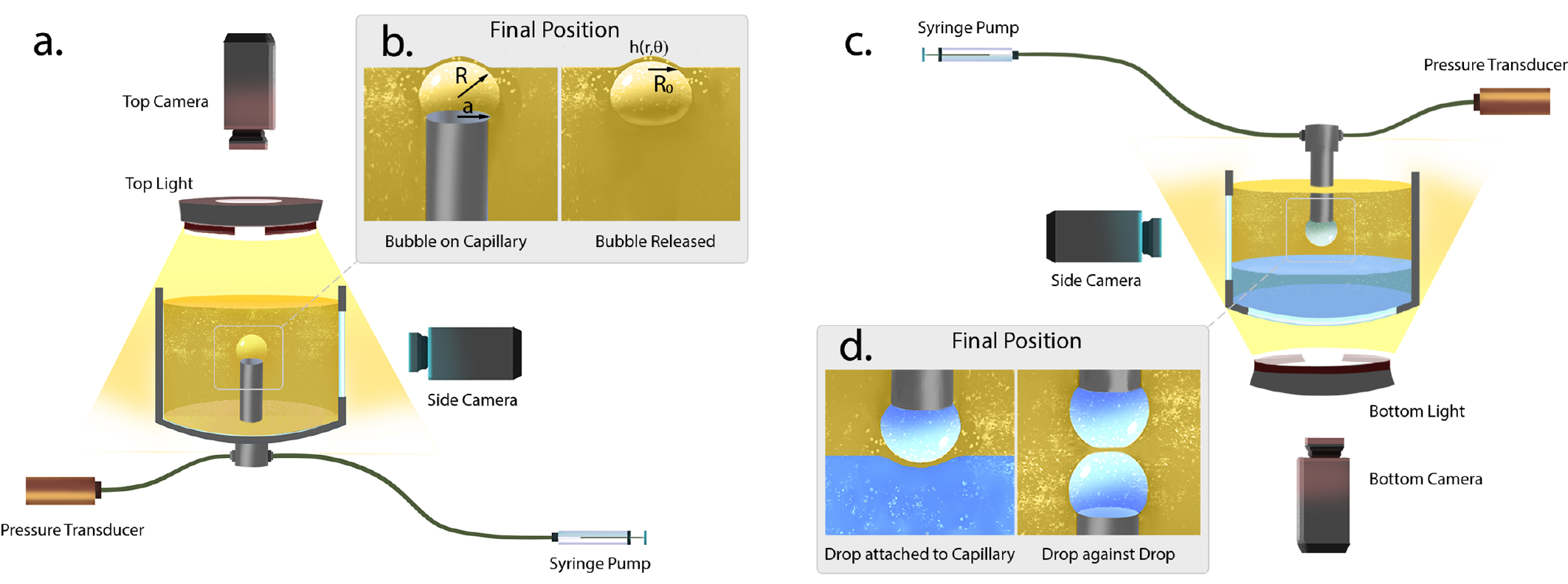} \caption{Schematic of single bubble and drop setups. ({\bf a.}) A schematic of a single bubble setup with the labeled components. ({\bf b.})  The final positions of the bubble for different variations of the setup. These include the cases where the bubble is attached to the capillary (left) and when it is released from the capillary (right). Here, $R$ is the radius of curvature of the undeformed bubble, $h(r,\theta)$ is the film thickness as a function of the radial position $r$ and angular position $\theta$, and $R_0$ is the radial extent of the film visible during thin film interferometry measurements. ({\bf c.}) A schematic of a single drop setup with the labeled components. Here, the drop is denser than the ambient liquid. ({\bf d.}) The final positions of the drop for different variations of the setup. These include the cases where the drop is attached to the capillary (left) and when interactions occur between two drops attached to capillaries (right).}\label{fig:ExperimentalSetup}
\end{figure*}

Detailed investigations into the interfacial rheological properties were also spawned in part as a result of Plateau's studies, where he claimed (though incorrectly in that setting \cite{fuller2012complex}) the existence of an interfacial viscosity through his description of the damping of a needle oscillating on the surface of an aqueous surfactant solution \cite{marangoni1972principle}. Interestingly, it was single drop experiment results -- the Stokes motion of liquid droplets -- that led Boussinesq to formulate the first mathematical description of interfacial viscoelasticity in 1913 \cite{boussinesq1913existence,fuller2012complex}. These results were later generalized for a Newtonian interface by Scriven in 1959 \cite{scriven1960dynamics}. In the subsequent years, techniques employing the controlled dynamic deformation of bubbles and drops were developed as a means to measure interfacial properties. Notably, efforts by Lunkenheimer and others in 1970's formed the basis for  oscillating bubble/drop rheometers \cite{lunkenheimer1984investigation}, while efforts by Darsh Wasan and others formed the basis for expanding and contracting bubble/drop rheometers \cite{nagarajan1993measurement}.

The stability of single bubbles and drops has attracted the attention of researchers since the 1800's due its fundamental \cite{rayleigh1882xx} and practical importance \cite{exerowa2018foam, clayton1923theory}. Some of the initial single bubble/drop experiments, notably by Lord Rayleigh \cite{rayleigh1882xx} and Geoffrey Ingram Taylor \cite{wilson1925bursting}, probed the bubble/drop stability against breakup in electrical fields. Single bubble experiments, notably by James Dewar, were also commonly used to study the important phenomenon of diffusion across liquid interfaces \cite{dewar1919soap}. One the earliest reported schematics that we can now identify as a single bubble/drop setup for studying bubble/drop stability can be observed in Fig. 54 of Charles Vernon Boys' popular book \textit{Soap-Bubbles and the forces which mould them} \cite{boys1890soap}. Early investigations into thin film stability by Derjaguin and  Kussakov that predated the development of the famed DLVO theory were also performed with single drop experiments \cite{derjaguin1939anomalous}. More practical versions of single bubble/drop setups can be seen in the works of Rehbinder and Wenstrom \cite{rehbinder1930stabilisierende}, while a feature complete version of a single bubble setup consisting of an arrangement to form bubbles in a controlled way along with an interferometry setup for measuring the film thickness can be found in a work published by Stanley Mason in 1960 \cite{charles1960coalescence}.

\section{Methods}\label{sec:Methods}
\subsection{Single bubble and single drop setups}
Single bubble and single drop setups provide a convenient framework to study in detail the dynamics of bubble-bubble and drop-drop interactions. Such an understanding is crucial to predict and tune the various aspects of foams and emulsions, including their stability \cite{suja2018evaporation} and density \cite{frostad2016dynamic}.   

A typical schematic of a single bubble and single drop setup is shown in Fig. \ref{fig:ExperimentalSetup}. A single bubble setup (Fig.\ref{fig:ExperimentalSetup}.a) commonly consists of a chamber to contain the ambient liquid, a capillary, and a syringe pump to form the bubble. In many cases, a pressure transducer is also connected to the capillary for monitoring the bubble pressure. The bubble pressure data is useful for many purposes including determining coalescence events (see \bit{Section \ref{subsec:coalescencedistributions}}) and measuring the rheological response of the air-liquid interface (see \bit{Section \ref{subsec:interfacialrheology}}). The profile of the bubble is visualized by a side camera, while the spatiotemporal evolution of the ambient liquid between the bubble and the air-liquid interface is visualized by the top camera (see \bit{Section \ref{subsec:thinfilmreconstruction}}). Further details of the setup depend on the type of the single bubble experiment.   

The different types of single bubble experiments reported in literature can be broadly classified into three categories. Namely, bubble attached to a capillary interacting with a flat air-liquid interface \cite{frostad2016dynamic,suja2018evaporation,suja2020foam,chandran2016impact, kannan2018monoclonal, kannan2019linking, suja2020symmetry}, bubble released from a capillary interacting with an air-liquid interface \cite{allan1961approach, zhang2015partial, feng2016dynamics, krzan2003pulsation, menesses2019surfactant, sunol2010rise, sett2013gravitational}, and bubble attached to a capillary interacting with another bubble on a capillary \cite{paulsen2014coalescence}. The final position of the bubble for two of these variants is shown in Fig.\ref{fig:ExperimentalSetup}b. Each of the above single bubble variants has specific advantages. The first variant, where the bubble remains attached to the capillary, is very well suited for studying thin film dynamics using interferometry, as the bubble position can be accurately controlled. Further advantages include the ability to easily control the ascend velocity and the size of the bubble. The second variant, where the bubble is released from the capillary, offers a convenient framework to study the bounce dynamics \cite{feng2016dynamics,sunol2010rise, krzan2003pulsation} of a freely rising bubble at an air-liquid interface. Finally, the third variant with two bubbles is very well suited to study bubble-bubble coalescence. These experiments are also more amenable to mathematical modelling due to the additional reflection symmetry in the physical configuration. The specific protocols corresponding to experiments in these different categories are described in the above references and are discussed to some extent in \bit{Section \ref{sec:FoamStability}}. For illustration, we outline below a protocol commonly followed for experiments in the first category. 

At the start of a single bubble experiment where the bubble remains attached to the capillary, fluctuations in the size of the bubble can be detected by monitoring the pressure inside the bubble. After establishing the size stability of the bubble, the experiment starts by moving the bubble at a fixed velocity towards the air-liquid interface from its initial to its predetermined final position (Fig.\ref{fig:ExperimentalSetup}b). In practice, for keeping the bubble in focus for interferometry measurements, the positioning of the bubble is accomplished by lowering the air-liquid interface towards the bubble by mechanically moving the chamber downwards. The final position of the bubble is usually selected such that it corresponds to the equilibrium position attained by a free bubble through the balance of buoyancy and capillary forces. The top camera records the spatiotemporal evolution of the film of liquid between the bubble and the air-liquid interface. As the film drains and its thickness becomes comparable to the wavelength of light, interference patterns are seen by the top camera (eg. see Fig.\ref{fig:Interferograms}). Finally, the experiment ends as the film ruptures and the bubble coalesces at some critical film thickness.  The coalescence time is accurately identified with the help of a pressure transducer. The details on analyzing the coalescence times and interference patterns are provided in the subsequent subsections. 

Single drop setups are broadly similar to single bubble setups. Since drops can either be lighter or heavier than the ambient liquid, single drop setups often have the capability to orient and move the drop in the direction of its natural motion (Fig.\ref{fig:ExperimentalSetup}c). As with the single bubble setup, there are three common variants of single bubble setups reported in the literature. Namely, drop attached to a capillary interacting with a flat liquid-liquid/solid interface \cite{bochner2017droplet,bluteau2017water}, drop released from a capillary interacting with a liquid-liquid/gas interface \cite{blanchette2006partial,mackay1963gravity, zhang2015partial,gillespie1956coalescence, aarts2008droplet}, and drop attached to a capillary interacting with another drop on a capillary \cite{frostad2016coalescence,paulsen2014coalescence,klaseboer2000film,vakarelski2010dynamic}. The final drop position for two of these variants is shown in Fig.\ref{fig:ExperimentalSetup}d. The protocols and advantages of the single drop variants more or less mirror those of single bubble setups and are discussed in context in \bit{Section \ref{sec:EmulsionStability}}.

\subsection{Coalescence time distributions}\label{subsec:coalescencedistributions}

A major observable from single bubble/drop experiments is the time it takes for a bubble or a drop to coalesce against a suitable air/liquid-liquid interface.  This quantity is commonly referred to as the coalescence time. The coalescence time of single bubbles and drops is physically correlated to the stability of foams \cite{suja2018evaporation, suja2020foam} and emulsions \cite{kamkar2020polymeric}, respectively, and provides a convenient way to predict and rank foam and emulsion stability. Unfortunately, directly using coalescence times to assess foam or emulsion stability might not give the intended results due to following three reasons. Firstly, single bubble/drop coalescence times are inherently stochastic \cite{suja2018evaporation,zheng1983laboratory,vitry2019controlling}. Secondly, the presence of coalescence modifiers such as antifoams or demulsifiers can lead to very large variations in the measured coalescence times \cite{suja2020foam}. Thirdly, coalescence times may have temporal trends \cite{poulain2018ageing, kannan2018monoclonal}.  Hence, rigorously predicting and ranking foam and emulsion stability from single bubble/drop measurements, requires careful statistical analysis. One such possibility is the use of coalescence time distributions.
\begin{figure}
    \centering
    \includegraphics[width=\columnwidth]{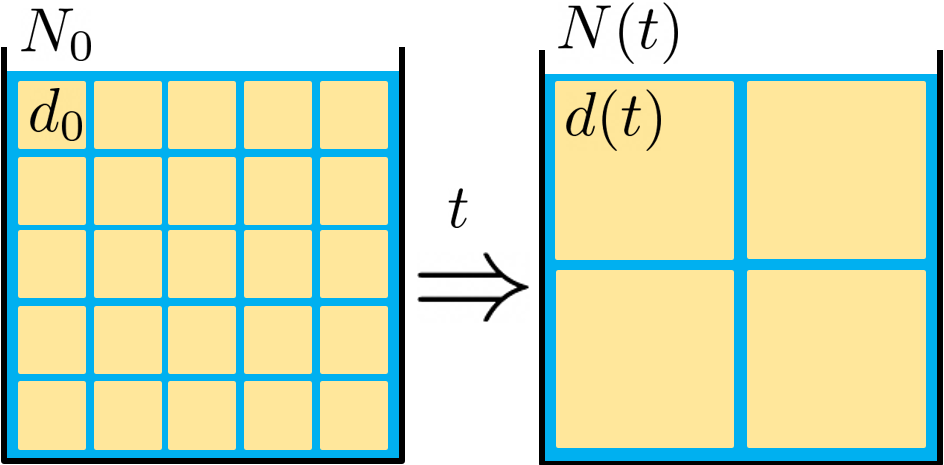}
    \caption{Schematic of an emulsion as a stack of mono-disperse cubic cells that decease in number and increase in size as time passes.}
    \label{fig:jt_emulsion}
\end{figure}

To understand the fundamental concept of coalescence time distributions and its relation to foam/emulsion life time, let us analyze single bubble/drop experiments from a more analytical point of view. For illustrative purposes we will use a simple emulsion model \cite{Kabalnov1998coalescence,Deminiere1999cell}. Consider an emulsion as a stack of $N_0$ mono-disperse cubic cells of size $d_0$ (Fig. \ref{fig:jt_emulsion}). Without changing the volume of the continuous and disperse phases, let's assume that coalescence events take place in such a way that, as time passes, the emulsion becomes a stack of $N(t)<N_0$ mono-disperse cells of size $d(t)>d_0$. As coalescence is a random process, it is reasonable to assume that the lifetime of a specific thin film separating two cells, $\tau_{tf}$, should be inversely proportional to the area of the thin film ($A$) and the probability of coalescence per unit area and time ($f$), i.e.
\begin{align}
\tau_{tf}\approx \frac{1}{Af},
\label{jt_lifetime}
\end{align}
From Eq. \ref{jt_lifetime}, the number of cells must verify
\begin{align}
\frac{dN(t)}{dt}=-f A_t(t),
\label{jt_dndt}
\end{align}
where $A_t(t)$ is the total surface area. Since we are considering a cubic cell system, Eq. \ref{jt_dndt} can be rewritten as
\begin{align}
\frac{dN(t)}{dt}= -3f N(t)d^2(t).
\label{jt_dndt2}
\end{align}
As above-mentioned, the volume of the disperse phase is constant, so that
\begin{align}
N(t)d^3(t)=k,
\label{jt_k}
\end{align}
being $k$ a constant. Combining Eqs. \ref{jt_dndt2} and \ref{jt_k} and integrating with respect time, we obtain the following relation (see \bit{Supplementary Materials} for details),
\begin{align}
\frac{1}{d_0^2}-\frac{1}{d^2(t)}&=2f t.
\label{jt_d02}
\end{align}

\begin{figure*}[!th]
\includegraphics[width=\linewidth]{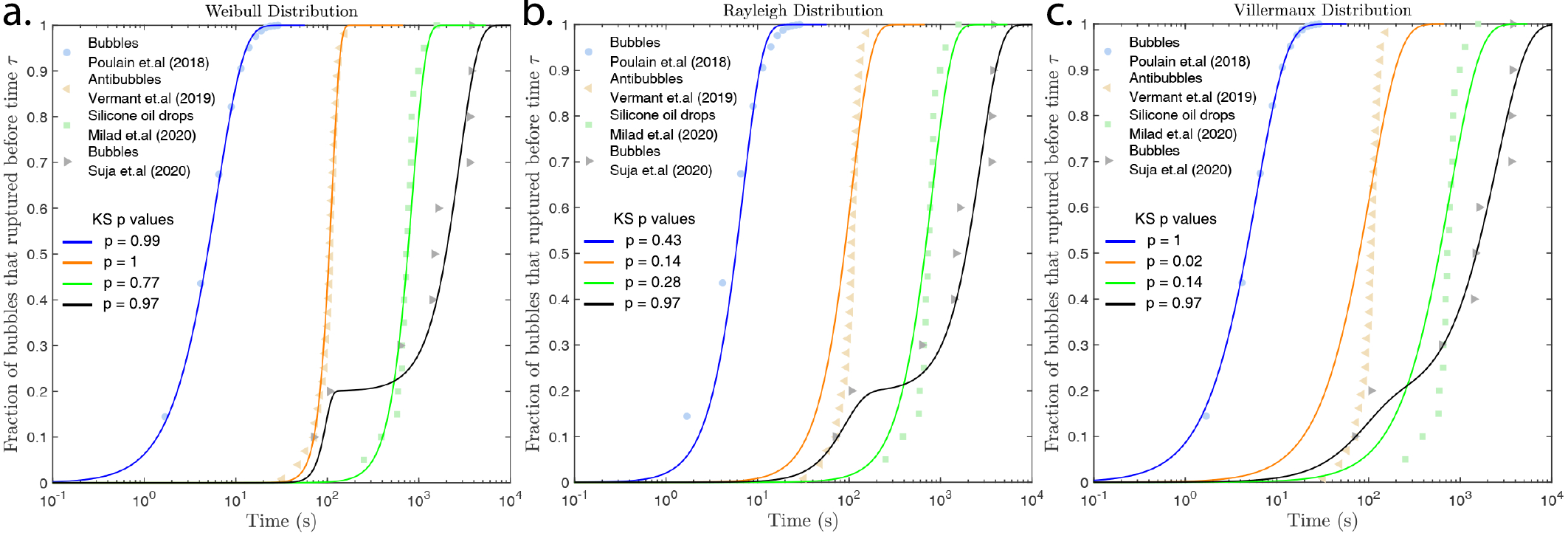} \caption{Comparison of three different probability distributions in capturing the distribution of coalescence times measured in four different systems: \protect\includegraphics[height=0.22 cm]{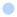} Bubbles in deionized water \cite{poulain2018ageing}, \protect\includegraphics[height=0.22 cm]{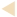} Antibubbles in a 10\% mixture of glycerol in water \cite{vitry2019controlling}, \protect\includegraphics[height=0.22 cm]{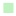} Silicone oil drops in an aqueous polymer mixture \cite{kamkar2020polymeric}, and \protect\includegraphics[height=0.22 cm]{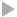} Bubbles in lubricants with antifoam \cite{suja2020foam}. The Kolmogrov-Smirnov (KS) $p$ values are indicated in the legend. ({\bf a.}) The two parameter Weibull distribution fit to the experimental data using the Maximum Likelihood Estimaters (MLE) of the scale and shape parameter. ({\bf b.})  The one parameter Rayleigh distribution fit to the experimental data using the MLE of the scale parameter.  ({\bf c.}) The one parameter Villermaux distribution fit to the experimental data using the MLE of the scale parameter.  The parameters and $R^2$ values of the plotted distributions are available in \bit{Supplementary Materials}.}\label{fig:CCCComparison}
\end{figure*}

In spite of the simplicity of this model, the linear relation between $1/d^2(t)$ and $t$ has been experimentally observed \cite{Deminiere1999cell}. The foam/emulsion lifetime, $\tau$, can be obtained from Eq. \ref{jt_d02} by imposing $1/d^2(\tau)=0$ as,
\begin{align}\label{eq.jt_lifetimeFreqLink}
\tau=\frac{1}{2d_0^2f}.
\end{align}
As explained in \bit{Section \ref{sec:Introduction}}, bulk foam/emulsion experiments have a high degree of complexity due to many body interactions, effects of advection, and the presence of plateau borders. Single bubble/drop experiments simplify the experimental approach to the problem by allowing for the systematic measurement of quantities such as the critical film thickness, drainage rates and the  interaction forces. In addition, the measurement of bulk emulsion/foam stability is also feasible, since a convenient number of single bubble/drop experiments allows one to construct a coalescence time distribution (Fig.\ref{fig:CCCComparison}). This statistical distribution characterizes the frequency of coalescence events $f$ previously described and is therefore directly related to the emulsion/foam lifetime as shown in Eq.\ref{eq.jt_lifetimeFreqLink}. 

An early use of coalescence time distributions can be seen in a work by Stanley Mason and co-workers, where they used a Gaussian distribution to capture the stability of surfactant-laden bubbles \cite{charles1960coalescence}. Since then, a number of statistical distributions including the Weibull \cite{tobin2011public}, Rayleigh \cite{suja2018evaporation, suja2020foam, zheng1983laboratory}, and custom distributions \cite{ghosh2004coalescence, poulain2018ageing, lhuissier2012bursting} have been used to capture life time of bubbles \cite{ghosh2004coalescence, suja2018evaporation, suja2020foam, zheng1983laboratory, poulain2018ageing, lhuissier2012bursting,tobin2011public}, drops \cite{ghosh2002analysis,kamkar2020polymeric}, and antibubbles  \cite{vitry2019controlling}. Despite the variety of distributions reported in the literature, interestingly, most of the commonly used coalescence time distribution can be shown to be a form of the Weibull distribution.

The Weibull distribution is a two parameter continuous distribution of positive random variables that is commonly used to describe the failure time of physical entities \cite{sornette2006critical}. The distribution has the following cumulative distribution function,
\begin{equation}\label{eq:WeibullDistribution}
    P_w(t;\lambda;k) = 1 - e^{-(t/\lambda)^k}.
\end{equation}
Here $t,\lambda,k$ are positive quantities, with $t$ denoting the measured coalescence time, $\lambda$ dictating the scale of the distribution, and $k$ dictating the shape of the distribution. The values for $\lambda$ and $k$ are usually obtained from their maximum likelihood estimators (see \bit{Supplementary Materials}). Two of the other commonly used  distributions, the Rayleigh distribution and the distribution reported by Villermaux and coworkers (hereon the Villermaux distribution) are variants of Weibull distribution with different values for $\lambda$ and $k$. 

The Rayleigh distribution is obtained by setting $\lambda = \sqrt{2}\sigma$ and $k=2$. The corresponding cumulative distribution function becomes,
\begin{equation}\label{eq:RayleighDistribution}
    P_r(t;\sigma) =  1 - e^{-t^2/(2\sigma^2)}.
\end{equation}
Here, $\sigma$ is the scale parameter of the distribution and is usually obtained from the maximum likelihood estimation method (see \bit{Supplementary Materials}).

The Villermaux distribution is obtained by setting $\lambda = \tau_0$ and $k=4/3$. The corresponding cumulative distribution function becomes,
\begin{equation}
    P_v(t;\sigma) =   1 - e^{-(t/\tau_0)^{4/3}}.
\end{equation}
Here, $\tau_0$ is the scale parameter of the distribution and is obtained from physical considerations as
\begin{equation}\label{eq.villemauxtau}
    \tau_0 = \left(\frac{4}{3\epsilon}\right)^{3/4} \left(\frac{R}{l_c}\right)^{1/2} \frac{\mu l_c}{\sigma_{\alpha\beta}},
\end{equation}
where $R$ is the radius of the bubble, $l_c$ is the capillary length, $\mu$ is the dynamic viscosity of the ambient fluid, $\sigma_{\alpha\beta}$ is the surface tension, and $\epsilon$ is an \textit{ad hoc} parameter that characterizes the bubble rupture efficiency. 

In figure \ref{fig:CCCComparison}, we compare the above three probability distributions in describing the distribution of coalescence times measured in four different systems - bubbles in deionized water \cite{poulain2018ageing}, antibubbles in a 10\% mixture of glycerol in water \cite{vitry2019controlling}, silicone oil drops in an aqueous polymer mixture \cite{kamkar2020polymeric}, and bubbles in lubricants with antifoam \cite{suja2020foam}. The scale and shape (when applicable) were obtained from the maximum likelihood estimators, while the mixture ratios (when applicable) were obtained using the expectation-maximization algorithm (see \bit{Supplementary Materials}). The Kolmogrov-Smirnov (KS) $p$ values of the obtained distributions are indicated in the figure legend, while the parameters and $R^2$ values of the distributions are available in \bit{Supplementary Materials}. It is worth noting that the obtained shape parameter is greater than $1$ in all the cases, which in the context of Weibull distribution physically implies that coalescence is more likely to happen as time proceeds.

As expected, we observe that the generic two parameter Weibull distribution best describes all the experimental data. This observation is supported by the high values of the KS $p$ metric. Despite the high fit fidelity, the presence of two parameters leads to practical difficulties such as ranking the coalescence stability and using the expectation-maximization algorithm \cite{moon1996expectation} for robustly determining the different distributions in the experimental data (see \bit{Supplementary Materials}).  The one parameter Rayleigh distribution is observed to broadly describe all the tested experiments. This observation is supported by the moderate values of the KS $p$ and $R^2$ metrics. Despite having only empirical evidence for its suitability \cite{zheng1983laboratory,suja2018evaporation,suja2020foam}, the Rayleigh distribution is a very convenient method for ranking the coalescence stability of diverse systems and for robustly representing the mixture distributions in the experimental data. The one parameter ($\epsilon$ in Eq.\ref{eq.villemauxtau} is a free parameter) Villermaux distribution is observed to very accurately describe the coalescence time distributions involving bubbles, while it appears to be relatively inaccurate when it comes to antibubbles and drops. This observation is supported by the high values of the KS $p$ and $R^2$ metrics for bubbles and relatively low values of the same metrics for the other systems. This is not surprising as the Villermaux distribution was derived for bubbles based on physical considerations \cite{lhuissier2012bursting}, and is a very convenient choice for describing bubble lifetimes and ranking bubble stability. It would be interesting for future studies to develop distributions utilizing physical arguments that can capture the experimental trends in antibubbles and drops. Particularly, these new distributions should be able to physically account for the variance in the measured coalescence times that appears to scale inversely with the dispersed phase viscosity.

\subsection{Thin film profile reconstruction}\label{subsec:thinfilmreconstruction} 

\begin{figure*}[!th]
\includegraphics[width=\linewidth]{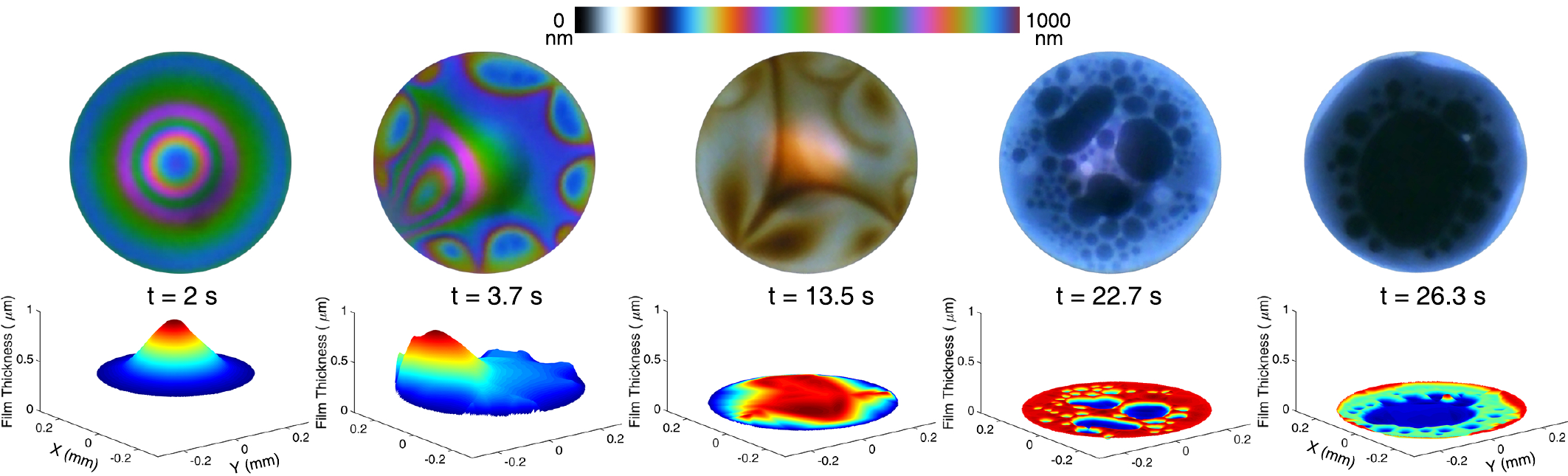} \caption{A sequence of interferograms observed over a bubble in a $10~$mM CTAB solution and the corresponding physical film thickness profiles reconstructed utilizing Eq.\ref{eq.SpectralIntensities}  (see \bit{Supplementary Materials}). The corresponding time stamps  are indicated below the interferograms with $t=0$ indicating the time at which the bubble encounters the flat air-water interface. The theoretical color map used for the reconstruction of the film profiles is shown at the top. Note that in the reconstructed film profiles, the blue and red colors correspond respectively to the minimum and maximum film thickness for each plot, and that the absolute film thickness can be inferred from the relative position of film profile about the $z$ axis.}\label{fig:Interferograms}
\end{figure*}

Measuring the spatiotemporal evolution of the film thickness is crucial for obtaining mechanistic insights into foam and emulsion stability. The film thickness data can be readily obtained from single bubble/drop experiments through the integration of a thin film interferometry apparatus.

Commercial and custom made interferometry apparatuses have been widely used in the literature for thin film thickness measurements. Commercial interferometers reported in the literature include those produced by Filmetrics \cite{bhamla2016instability,hermans2015lung}, Zygo \cite{shukla2006non,greco1994measuring}, and Horiba \cite{habibi2007langmuir}, among others. These interferometers are particularly efficient at high frequency automated thickness measurements at low spatial resolutions, often restricted to a single point. In many scenarios involving foam and emulsion films, it is necessary to measure the film thickness at both high temporal and spatial resolutions. This has motivated a number of researchers to perform studies with custom built thin film interferometry apparatuses \cite{bluteau2017water,zhang2015domain,zhang2016nanoscopic,allan1961approach,mackay1963gravity,lv2012spatial,vannoni2013measuring,beltramo2016simple,beltramo2016millimeter,kralchevsky1990interaction,nikolov1989ordered,hendrix2012spatiotemporal,frostad2016dynamic,bochner2017droplet,kannan2018monoclonal,kannan2019linking,suja2018evaporation,suja2020foam,czakaj2020viscoelastic,rodriguez2019evaporation,rabiah2019influence,joye1994asymmetric,nierstrasz1999marginal,lee2014surfactant,rabiah2020understanding}. 

As shown in Fig.\ref{fig:ExperimentalSetup}, these custom built interferometers generally consists of a light source, an optical train, and a photo-detector. Common light sources include laser based monochromatic \cite{beltramo2016simple,beltramo2016millimeter,hendrix2012spatiotemporal,nikolov1989ordered}, optically filtered LED based monochromatic \cite{bluteau2017water}, and LED or halogen based broadband sources \cite{kitagawa2013thin,frostad2016dynamic,bochner2017droplet,kannan2018monoclonal,kannan2019linking,suja2018evaporation,suja2020foam,czakaj2020viscoelastic,rodriguez2019evaporation,rabiah2019influence}. Common optical components include a lens assembly usually in the form of microscope objectives \cite{nikolov1989ordered,beltramo2016simple,bluteau2017water} or telecentric lenses \cite{frostad2016dynamic,kannan2018monoclonal,suja2018evaporation,rodriguez2019evaporation}, and optical filters \cite{mackay1963gravity,frostad2016dynamic,kitagawa2013thin,bluteau2017water}. Routinely used photo-detectors to image the interferograms include cine \cite{mackay1963gravity}, CCD \cite{kitagawa2013thin}, and CMOS \cite{frostad2016dynamic,hendrix2012spatiotemporal} cameras. The obtained interferograms are then decoded to obtain the underlying film thickness. This is accomplished by utilizing results from the theory of thin film interference. 

The theory of thin film interference was formalized in the early 19\textsuperscript{th} century by Fresnel, and has since been discussed by many researchers in the context of measuring the thickness of thin films such as for bubbles \cite{sheludko1967thin} and tear films \cite{doane1989instrument}, and for surface profiling \cite{blodgett1952step}. Here we will briefly develop two formulations relevant for the custom setups reported in the literature.  For this let us consider a beam of light having intensity $I_0(\lambda)$ and wavelength $\lambda$ incident on a thin liquid film of thickness $h$ and refractive index $n_2$. The film is bounded on top and bottom by media having refractive indices $n_1$ and $n_3$, respectively. Before proceeding we will assume that the angle of incidence is small and the film is non-dispersive and non-absorbing (see \bit{Supplementary Materials} for complete derivations).

\subsubsection*{Two reflections}
In the first formulation, considering the first two reflections and neglecting contributions from higher order reflections, we obtain \cite{kitagawa2013thin,hecht2017optics}, 
\begin{align}\label{eq.LightIntensities}
   \frac{ I(\lambda,h)}{I_0(\lambda)} &=   \mathfrak{R}_1 +\mathfrak{R}_2(1 - \mathfrak{R}_1)^2   + 2\sqrt{\mathfrak{R}_1\mathfrak{R}_2(1 - \mathfrak{R}_1)^2}\cos\left(\phi  \right),\\
     \phi &= \frac{4\pi n_2 h}{\lambda} + \pi (\mathds{1}(n_2>n_1)) + \pi (\mathds{1}(n_3>n_2)).
\end{align}
Here, $\phi$ is the phase difference between the interfering reflected beams and $\mathds{1}$ is the indicator function that captures the phase shift of $\pi$ radians that occurs when light reflects off a medium with a higher refractive index. $\mathfrak{R}_1$ and $\mathfrak{R}_2$ are the power (intensity) reflectivity coefficients obtained from the Fresnel equations evaluated for normal incidence, and are given by,
\begin{align}
    \mathfrak{R}_1 = r_{12} = (-r_{21})^2 = \left(\frac{n_1 - n_2}{n_1 + n_2}\right)^2,\\
    \mathfrak{R}_2 = r_{23} = (-r_{32})^2 = \left(\frac{n_2 - n_3}{n_2 + n_3}\right)^2,
\end{align}
where $r_{xy}$ are the Fresnel amplitude reflectivity coefficients. 

Finally, the intensity perceived by the $i^{th}$ channel of a pixel $P$ in a camera as a function of the film thickness can be computed as, 
\begin{equation}\label{eq.SpectralIntensities}
    P_i (h) = \int_{\lambda_0}^{\lambda_f}  I(\lambda,h) I_r(\lambda) S_i(\lambda) d\lambda.
\end{equation}
Here, $I_r(\lambda)$ is the spectral response of the optical components in the system and $S_i(\lambda)$ is the spectral sensitivity of the $i^{th}$ channel of a pixel. $\lambda_0$ and $\lambda_f$ are the lowest and the largest wavelengths with non-zero intensities that contribute to the signal in the photo-detector. 

 Utilizing Eq.\ref{eq.SpectralIntensities}, we can invert the interferograms to recover the thickness of the thin film.  For example, Figure \ref{fig:Interferograms} shows the reconstructed film thickness profiles using Eq.\ref{eq.SpectralIntensities} from interferograms obtained over a bubble in a $10~$mM CTAB solution. The reconstruction procedure involving correlating the colors in the color map (Fig.\ref{fig:Interferograms} inset) to the colors in the interferogram is detailed in Frostad et al.\cite{frostad2016dynamic}.
 
\subsubsection*{Infinite reflections}
 In the second formulation, considering all the reflected waves, we obtain,
 \begin{align}\label{eq.IntensityInfinite}
     \frac{I (\lambda,h)}{I_0 (\lambda)} = 1 - \frac{(1-r_{12}^2) (1 - r_{23}^2)}{ (1 - r_{23}r_{21})^2 + 4r_{23}r_{21}\sin^2 \beta}.
 \end{align}
 
\noindent Introducing,
 \begin{equation*}
    \Delta (\lambda) = \frac{I (\lambda,h) - I_{min}(\lambda)}{I_{max}(\lambda) - I_{min}(\lambda)},
\end{equation*}
we obtain the following expression for the film thickness \cite{sheludko1967thin,karakashev2008effect},
\begin{equation}\label{eq.ThicknessInfinite}
    h = \frac{\lambda}{2\pi n_2}\left(l\pi\pm\arcsin  \sqrt{\frac{\Delta (1 - r_{23}r_{21})^2}{(1 - r_{23}r_{21})^2 +  4r_{23}r_{21} (1 - \Delta)}}\right).
\end{equation}
Here, $\beta = 2\pi n_2 h/\lambda$ and $l$ is a whole number that denotes the order of interference. The common protocols for determining $l$ as well as the details related to using Eq. \ref{eq.ThicknessInfinite} to recover film thickness profiles are available in the literature \cite{karakashev2008effect,sett2013gravitational,beltramo2016simple, zhang2015domain}. 

Eq.\ref{eq.ThicknessInfinite} is a convenient choice when experiments are performed using monochromatic light sources in situations where the order of interference ($l$) can be easily inferred. On the other hand, when broadband light sources are used or when it is difficult to determine $l$ (eg. for films that do not thin below a few hundred nanometers), Eq.\ref{eq.SpectralIntensities} is a convenient choice.  Note that the truncation error in using Eq.\ref{eq.SpectralIntensities} is $\mathcal{O}(r_{xy}^8)$, which in almost all practical cases is negligible (see \bit{Supplementary Materials}). 

We will conclude this section by touching upon techniques apart from traditional interferometry that have been used to measure the spatiotemporal profiles of thin liquid films. Planar laser induced fluorescence (PLIF) is a common technique used to visualize liquid films \cite{aarts2008droplet,dong2019laser,bordoloi2012effect}, and is particularly well suited to study thick (sub-millimeter scale) films. Hyperspectral interferometry is a technique in development that has improved robustness against imaging noise \cite{suja2020hyperspectral}.   Digitally holography is a promising technique that has recently been employed visualize bubbles \cite{mandracchia2019quantitative}. The large measurement range and high spatiotemporal resolution of digital holography could make this the technique of choice for future studies involving thin films.  

\subsection{Interfacial rheological properties}\label{subsec:interfacialrheology}

The stability of foams and emulsions is significantly influenced by the rheological properties at the fluid-fluid interface \cite{frostad2016coalescence,kannan2018monoclonal, kannan2019linking, lin2016interfacial, lin2018influence, kannan2020surfactant, rodriguez2020asphaltene, nagel2017drop, jaensson2018tensiometry, kotula2015regular, fuller2012complex, wilde2004proteins}. These so-called ``complex'' or ``non-Newtonian'' interfaces arise due to the presence of adsorbed surface active species, which can laterally interact and form microscopic networks that allow the interface to support both shear and normal stresses \cite{fuller2012complex, hermans2015lung, sanchez2018dynamic, scriven1960dynamics}. Unlike simple interfaces,  these stresses can occur in the absence of a finite curvature or gradients in surface tension \cite{fuller2012complex}. 

Complex interfaces exhibit a viscoelastic response to surface deformations. In other words, the resulting stress exhibits both a strain-dependent (elastic) and a strain rate- dependent (viscous) response \cite{fuller2012complex, sanchez2018dynamic, berry2015measurement, carvajal2011mechanics, nagel2017drop, danov2015capillary}. Non-Newtonian interfaces are thus described via rheological constitutive equations that take into account the time-, position-, and velocity-dependent nature of their viscoelastic properties \cite{fuller2012complex, sanchez2018dynamic, berry2015measurement, carvajal2011mechanics, nagel2017drop, danov2015capillary}.

For a general viscoelastic liquid interface, total interfacial stress tensor $\boldsymbol{\sigma}$ can be expressed as \cite{jaensson2018tensiometry, hermans2015lung},
\begin{equation}\label{eq.InterfacialStress}
    \boldsymbol{\sigma}  = \sigma_{\alpha \beta}(\Gamma) \boldsymbol{I}_s + \boldsymbol{\sigma}_e.
\end{equation}
Here, $\sigma_{\alpha \beta}(\Gamma)$ is the static equilibrium value of the interfacial tension between phases $\alpha$ and $\beta$ as a function of surface species concentration ($\Gamma$) and  $\boldsymbol{I}_s$ is the second order surface identity tensor. The interfacial tension is a scalar thermodynamic quantity that provides a measure of the work required to increase the surface area of an interface \cite{leal2007advanced, berg2012introduction}.  $\boldsymbol{\sigma}_e$ is the extra rheological stress arising from viscous and elastic deformations. This stress is represented as a rank-2 tensor that describes how in-plane stresses propagate along each of the interfacial coordinate directions. In its most general form, this tensor is non-isotropic \cite{rodriguez2020asphaltene, nagel2017drop, danov2015capillary}. The remainder of this article will focus on displacements and stresses that exist purely in the tangential direction; for a discussion on bending and normal stresses, the reader is directed to references \cite{murphy1966thermodynamics, leermakers2006symmetric, evans1994hidden, helfrich1973elastic, helfrich1989bending, gekle2017theory}.

A general viscoelastic interface will have both a viscous and an elastic contribution to the extra stress, thus requiring an appropriate viscoelastic model to capture the combined contribution of viscous and elastic deformations. Depending on the nature of the interface and the deformation, a number of relations are available in the literature for calculating the extra rheological stresses. For example, for a purely viscous Newtonian interface, $\boldsymbol{\sigma}_e$ can be obtained from the Boussinesq-Scriven equation as \cite{scriven1960dynamics, lopez1998direct}
\begin{equation}\label{eq.InterfacialDeformationStressBS},
    \boldsymbol{\sigma}_e = \left[\left(k_s -\eta_s\right)\nabla_s \cdot \boldsymbol{v}_s\right]\boldsymbol{I}_s + 2\eta_s \boldsymbol{\mathds{D}_s},
\end{equation}
where $\nabla_s$ is the surface gradient operator, $\boldsymbol{v}_s$ is the surface velocity vector, and $\boldsymbol{\mathds{D}_s}$ is the surface rate of deformation tensor, equal to $\frac{1}{2}\left(\nabla_s  \boldsymbol{v}_s + \left( \nabla_s  \boldsymbol{v}_s \right) ^T \right)$. Eq. \ref{eq.InterfacialDeformationStressBS} shows that $\boldsymbol{\sigma}_e$ depends on two material properties, namely the surface dilatational viscosity ($k_s$) and the surface shear viscosity ($\eta_s$). Much like their bulk fluid counterparts, complex interfaces can support stresses when subject to shearing deformations and are similarly characterized by a shear viscosity \cite{fuller2012complex, scriven1960dynamics}. However, unlike bulk liquids, which behave as incompressible fluids in normal operating conditions, fluid-fluid interfaces are able to change their surface area when subject to dilatational or compressional deformations, and subsequently are also characterized by an interfacial dilatational viscosity \cite{fuller2012complex, scriven1960dynamics}. 

For a purely elastic interface undergoing small deformations, $\boldsymbol{\sigma}_e$ can be obtained from infinitesimal strain theory as follows \cite{jaensson2018tensiometry}
\begin{equation}\label{eq.InterfacialDeformationStressIST},
    \boldsymbol{\sigma}_e = \left[\left(K_s -G_s\right)\nabla_s \cdot \boldsymbol{u}_s\right]\boldsymbol{I}_s + 2G_s \boldsymbol{\mathds{U}_s},
\end{equation}
where $\boldsymbol{u}_s$ is the surface displacement vector and $\boldsymbol{\mathds{U}_s}$ is the surface deformation tensor, equal to $\frac{1}{2}\left(\nabla_s  \boldsymbol{u}_s + \left( \nabla_s  \boldsymbol{u}_s \right)^T \right)$. Analogous to the viscous contribution, $\boldsymbol{\sigma}_e$ is a function of the surface dilatational modulus ($K_s$) and the surface shear modulus ($G_s$). Alternatively, for larger deformations, the Neo-Hookean model for an elastic interface can be used \cite{pepicelli2017characterization,jaensson2018tensiometry}.

In this section, we will discuss how single bubble/drop setups can also be used as a convenient platform with little modification to measure the static interfacial stress and the surface dilatational properties.

\subsubsection{Static and Dynamic Interfacial Stress} \label{sec:IFT}

\subsubsection*{Pendant drop tensiometry}
\begin{figure*}[h]
\includegraphics[width=0.98\linewidth]{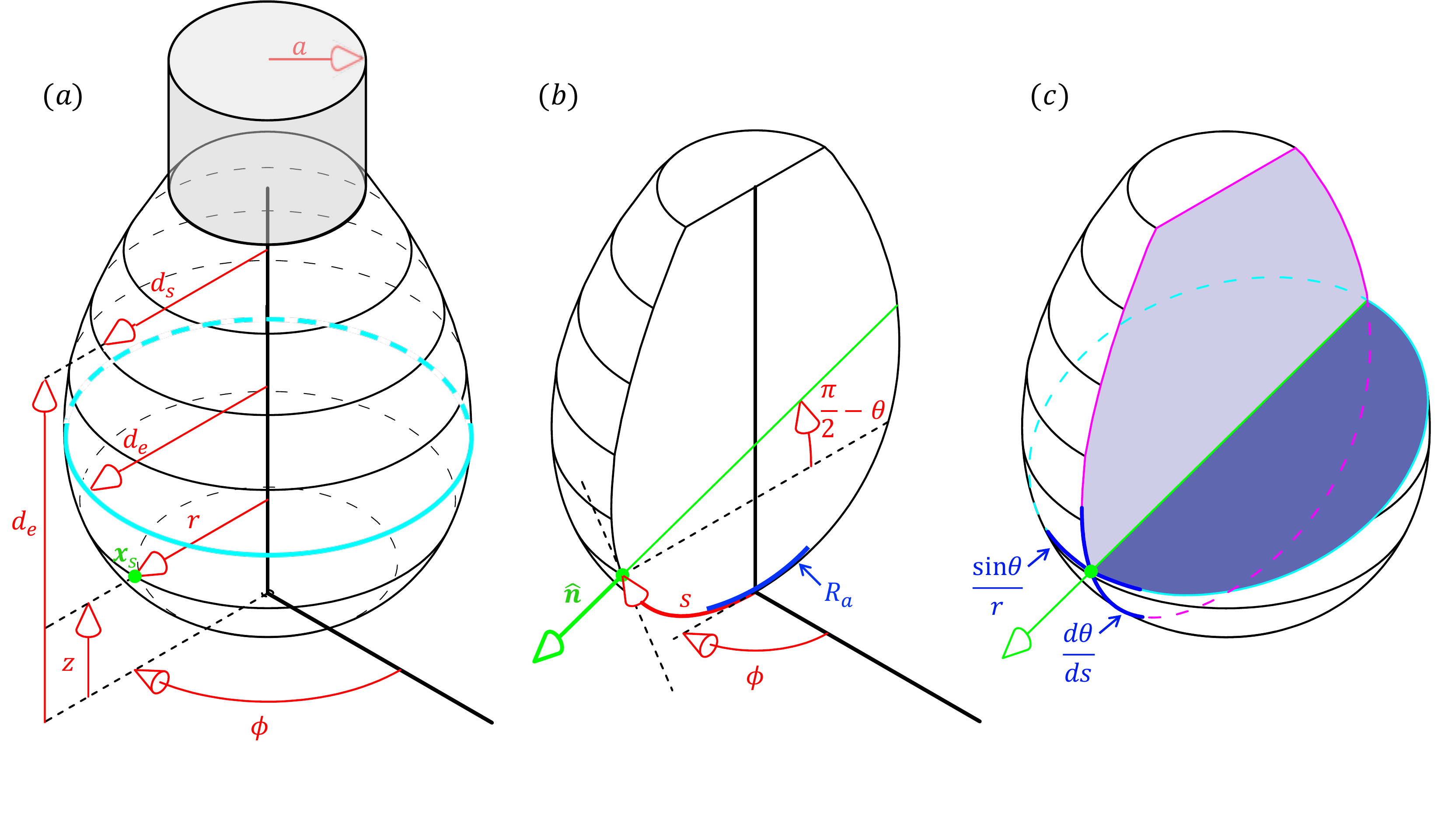} \caption{Schematic of an axisymmetric pendant drop suspended from a capillary with radius $a$. ({\bf a.}) A point on the surface of the drop, $\boldsymbol{x}_s$, can be uniquely described using a cylindrical coordinate system ($r-z-\phi$). $d_s$ and $d_e$ denote diameters (only half extend is shown) used in the method of the selected plane to calculate $\sigma_{\alpha\beta}$. ({\bf b.}) The coordinate system is projected onto the $s-\phi$ frame, where $s$ is the arc length measured from $z=0$ and $\phi$ is the azimuthal angle. $\theta$ is the angle formed between $s$ and the horizontal plane, such that $\pi/2 - \theta$ is the angle between the horizontal plane and the surface unit normal $\boldsymbol{\hat{n}}$. $R_a$ is the radius of curvature at the drop apex. ({\bf c.}) The principal curvatures $\frac{d\theta}{ds}$ and $\frac{\sin\theta}{r}$ are locally tangent to the $s$- and $\phi$-coordinates, respectively.}\label{fig:PendantDrop}
\end{figure*}

Pendant drop tensiometry is a common and robust technique for measuring the interfacial stress of liquid-liquid and liquid-air interfaces \cite{rotenberg1983determination, berry2015measurement}. In this method a pendant drop is formed on a capillary and its shape is iteratively fit to the theoretical shape obtained from the interfacial stress balance \cite{rotenberg1983determination, berry2015measurement}. 

For simple interfaces, the static and dynamic interfacial stress is solely determined by the interfacial tension, $\sigma_{\alpha \beta}(\Gamma)$, which is constant and isotropic along the drop’s surface \cite{berg2012introduction, leal2007advanced,deGennes2004capillarity, fuller2012complex}. A convenient way to measure $\sigma_{\alpha \beta}(\Gamma)$ is through the so called Axisymmetric Drop Shape Analysis (ADSA) method, whereby the interfacial tension is obtained by analyzing the shape of a stationary pendant drop in a gravitational field \cite{berry2015measurement, rotenberg1983determination, nagel2017drop}. The shape of the axisymmetric drop is set by a balance between gravitational deformation and surface tension restoration, represented by the dimensionless Bond number, $Bo = \Delta\rho g R_a^2/\sigma_{\alpha \beta}(\Gamma)$ \cite{berry2015measurement, rotenberg1983determination, nagel2017drop}. Here, $\Delta\rho$ is the difference in density between the drop and the bulk, $g$ is the gravitational acceleration, and $R_a$ is the radius of curvature at the drop apex. 

An axisymmetric pendant drop is depicted in Figure \ref{fig:PendantDrop}. Any point $\boldsymbol{x}_s$ at the surface of the drop can be described by a cylindrical coordinate system $(r-z-\phi)$ (Fig.\ref{fig:PendantDrop}a). This coordinate system can be projected onto the $s-\phi$ frame, where $s$ is equal to the arc length measured from $z=0$ and $\phi$ is the azimuthal angle (Fig.\ref{fig:PendantDrop}b) \cite{nagel2017drop, danov2015capillary}. In this coordinate system, $s$ and $\phi$ are both locally tangent to $\boldsymbol{x}_s$ and are related to the cylindrical coordinates via the following transformations,
\begin{align}
    \frac{dr}{ds}  &= \cos \theta, \label{eq.Transformation1}\\
    \frac{dz}{ds}  &= \sin \theta, \label{eq.Transformation2}
\end{align}

where $\theta$ is the meniscus slope angle (i.e. the angle formed between the horizontal plane and the drop interface). 

This coordinate transformation allows for a facilitated determination of the drop shape and is particularly advantageous for the computation of the interfacial stress tensor for non-isotropic complex interfaces, as outlined at the end of the section \cite{nagel2017drop}.

Using the coordinate transformations in Eqs. \ref{eq.Transformation1} and \ref{eq.Transformation2}, the isotropic value for the interfacial tension of an axisymmetric pendant drop or bubble is prescribed by the interfacial normal stress balance  \cite{rotenberg1983determination, berry2015measurement,alvarez2009non},
\begin{align}  \label{eq.YoungLaplace}
    \sigma_{\alpha\beta}(\Gamma) \left( \frac{d\theta}{ds} + \frac{\sin \theta}{r} \right) = \Delta P_a - \Delta\rho g z. 
\end{align}
Eqs. \ref{eq.Transformation1} -- \ref{eq.YoungLaplace} comprise a system of ordinary differential equations in curvilinear co-ordinates subject to the boundary condition,
\begin{equation} \label{eq.YoungLaplaceBC}
    r =0, \; z =0,\; \theta = 0\quad \text{at}\quad s = 0.
\end{equation}
Eq. \ref{eq.YoungLaplace} is the Young-Laplace equation, which relates the pressure jump across the interface, $\Delta P_a - \Delta\rho g z$, to the local principal meridional and parallel curvatures, $\frac{d\theta}{ds}$ and $\frac{\sin \theta}{r}$ (see Figure \ref{fig:PendantDrop}c). The pressure jump at any point along the interface can be obtained by adding the hydrostatic pressure contribution, $\Delta\rho g z$ (where $z=0$ corresponds to the position of the drop apex), to the pressure jump at the drop's apex, $\Delta P_a$ (determined via symmetry) \cite{rotenberg1983determination, berry2015measurement}. The sign of the gravitational term depends on whether the drop is buoyant or pendant. 

Eqs. \ref{eq.Transformation1} -- \ref{eq.YoungLaplace} are iteratively solved numerically for different values of $\sigma_{\alpha \beta}(\Gamma)$ until the solution converges to the experimentally obtained drop profile \cite{berry2015measurement, rotenberg1983determination}. The accuracy of the obtained surface tension scales inversely with the square of the Worthington number, defined as $Wo = Bo V_d/(2\pi a R_a^2)$, where $V_d$ is the volume of the drop and $a$ is the radius of the capillary \cite{berry2015measurement}. 

Despite being the most accurate way to recover $\sigma_{\alpha \beta}(\Gamma)$ from pendant drop images, numerically solving Eqs. \ref{eq.Transformation1} -- \ref{eq.YoungLaplace} is computationally intensive and time consuming. Alternatively, the method of the plane of inflection or the method of the selected plane may be used to determine the surface tension from pendant drop images \cite{andreas2002boundary}. The method of the selected plane is the more accurate method among the two, and involves recasting the bond number as $Bo_{de} = \Delta \rho g d_e^2/\sigma_{\alpha \beta}(\Gamma)$ and using the numerically tabulated values of $Bo_{de}$ as function of the drop shape parameter $S = d_s/d_e$ to recover  $\sigma_{\alpha \beta}(\Gamma)$ \cite{andreas2002boundary,juuza1997pendant,stauffer1965measurement}. Here, $d_e$ is the equatorial diameter of the drop and $d_s$ is the drop diameter at a height of $d_e$, both which are easily obtained through image processing (Fig. \ref{fig:PendantDrop}a). Additional details of the axisymmetric drop shape analysis, including a historical perspective and a discussion of variants of the technique such as the compound pendant drop technique \cite{neeson2014compound}, are available in Berry et al. \cite{berry2015measurement}.

Although complex interfaces are described by a position-dependent interfacial stress tensor $\boldsymbol{\sigma}$, there are cases where $\boldsymbol{\sigma}$ reduces to a constant scalar value, rendering traditional shape-fitting methods valid for the interfacial energy calculation. This can be achieved in complex interfaces if the interface remains undeformed so that deviatoric viscoelastic stresses are not present. Thus, in the absence of any surface deformations, the surface stress (Eq.\ref{eq.InterfacialStress}) on a stress-relaxed viscoelastic interface is isotropic and constant along the drop surface, and Eqs. \ref{eq.Transformation1} -- \ref{eq.YoungLaplaceBC} can be used. The validity of the ADSA method for complex interfaces can be verified by plotting the local mean curvature of the surface as a function of the height along the drop; if the slope is constant, then the interfacial stress is prescribed by a constant scalar \cite{nagel2017drop}. However, if a viscoelastic interface is deformed sufficiently, it will exhibit stress anisotropy. In such cases, Eq. \ref{eq.YoungLaplace} does not accurately describe the interface, and instead needs to be modified to account for the spatial and directional dependence of the surface stress. 

Danov and coworkers developed a method known as Capillary Meniscus Dynamometry (CMD), whereby the components of the anisotropic interfacial stress tensor are determined for axisymmetric drops/bubbles \cite{danov2015capillary}. Once again, the interfacial stress balances is simplified by projecting the coordinate system onto the $s - \phi$ frame. Since the $s$ and $\phi$ coordinates are locally tangent to the principal curvatures, the interfacial stress tensor $\boldsymbol{\sigma}$ can be diagonalized and its deviatoric components expressed in terms of a pair of principal stresses as, $\boldsymbol{\sigma} = \sigma_s\boldsymbol{\hat{e}}_s \boldsymbol{\hat{e}}_s + \sigma_{\phi}\boldsymbol{\hat{e}}_{\phi} \boldsymbol{\hat{e}}_{\phi}$, where $\boldsymbol{\hat{e}}_s$ and $\boldsymbol{\hat{e}}_{\phi}$ are the unit vectors  \cite{nagel2017drop, danov2015capillary}.

In the CMD method, the interface is split into small domains and the normal and tangential stress balances are applied locally \cite{danov2015capillary}
\begin{align}
    \sigma_s\frac{d\theta}{ds} + \sigma_{\phi}\frac{\sin \theta}{r}&= \Delta P_a - \Delta\rho g z , \label{eq.NormalStress}\\
    \sigma_{\phi} &= \frac{d \left(r \sigma_s \right)}{d r}. \label{eq.TangentiallStress}
\end{align}
 
The Laplace pressure at the apex, $\Delta P_a$, and the interface position are required as input parameters and can be determined from experimental measurements. This method was further developed by Nagel et al., who wrote a set of MATLAB routines that are available online under an open-source license \cite{nagel2017drop}. 

\begin{figure}[t]
\includegraphics[width=0.98\linewidth]{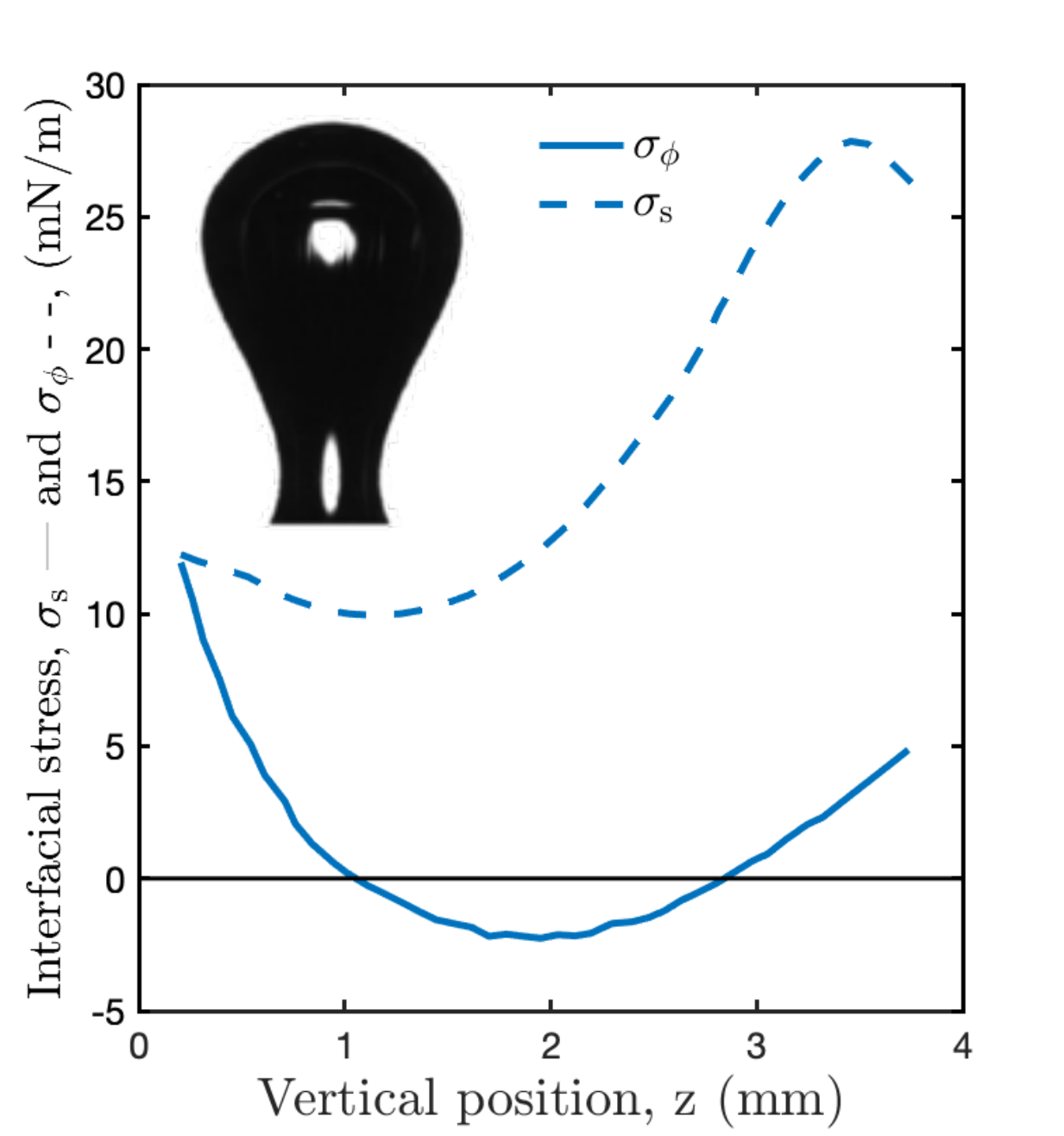} \caption{A buoyant bubble in a 0.005 wt\% aqueous HFBII solution is allowed to age for $320~$s and then compressed in a step-wise fashion. The Laplace pressure at the drop apex and the bubble shape are used to determine the anisotropic components of the surface stress, $\sigma_s$ and $\sigma_{\phi}$. These are plotted against the vertical coordinate $z$ measured from the drop apex at a time of $420~$s after bubble formation. Regions where $\sigma_s$ adopts negative values correspond to the appearance of wrinkles along the bubble surface. Results reproduced from Danov et al. \cite{danov2015capillary}.}\label{fig:Sigmas_CMD}
\end{figure}

Figure \ref{fig:Sigmas_CMD} reproduces results obtained by Danov et al. for a buoyant bubble in a 0.005 wt\% aqueous solution of the protein hydrophobin HFBII \cite{danov2015capillary}. The bubble was allowed to age for $320~$s and the Laplace pressure at the drop apex was measured as the bubble volume was reduced in a step-wise fashion. Upon compression, the bubble shape starts to deviate from the Young-Laplace equation. The two components of the interfacial stress, $\sigma_s$ and $\sigma_{\phi}$, are plotted against the vertical coordinate $z$ measured from the drop apex. Significant stress anisotropy is evidenced by the variation of $\sigma_s$ and $\sigma_{\phi}$ along the height of the bubble. The region where $\sigma_s$ adopts negative values corresponds to the appearance of wrinkles along the bubble surface.

In many applications, such as for characterizing  protein absorption \cite{kannan2019linking} and foam generation \cite{buzzacchi2006dynamic}, it is important to measure the dynamic value of surface stress. The axisymmetric drop shape analysis is convenient for this purpose, when the dynamic change of surface stress is slow (in the order of a few seconds) compared to time required to form a pendant drop and measure the stress. This is often the case when changes in the interfacial tension are brought about by the adsorption and rearrangement of large molecules at the interface \cite{kannan2019linking,rodriguez2020asphaltene, kotula2015regular}. On the other hand, when the change in interfacial tension is fast (on the order of milliseconds), such as with small molecule surfactants, it is necessary to resort to other techniques to measure the dynamic surface stress. 

\subsubsection*{Maximum bubble pressure method}
The maximum bubble pressure method is an appropriate technique for high frequency surface stress measurements \cite{fainerman1998maximum, buzzacchi2006dynamic,bendure1971dynamic} and can be conveniently performed in single bubble/drop setups. The method involves bubbling a fluid through a capillary and measuring the pressure as a function of the bubbling frequency. The capillary pressure reaches a maximum when the radius of curvature of the bubble/drop equals the radius of the capillary. Utilizing this information, an isotropic surface stress is recovered from the simplified Young-Laplace equation as below,   
\begin{equation}\label{eq.MaximumBubblePressure}
    \sigma_{\alpha\beta}(\Gamma(t_{max})) = \frac{aP_{cmax}}{2}f_c
\end{equation}
Here, $a$ is the radius of the capillary, $f_c$ is a shape correction factor that accounts for any deviation of the bubble from a spherical shape, $P_{cmax}$ is the maximum capillary pressure and $t_{max}$ is the time taken for attaining $P_{cmax}$ after forming a new bubble/drop (equivalently after releasing a bubble/drop). $P_{cmax}$ is obtained by subtracting the hydrostatic pressure ($P_h$) and the excess dynamic pressure $P_d$ from the pressure $P_t$ measured by the pressure transducer, i.e $P_{cmax} = P_t - P_h - P_d$. To obtain the surface stress as a function of the interface age, $t_{max}$ is varied by changing the bubbling frequency. Additional details about the technique including the calculation of $f_c$ and $P_d$, considerations for very high frequency measurements, and the dilatational contributions to the measured surface stress are available in Fainerman et al. \cite{fainerman1998maximum}.

\subsubsection*{Microscopic drops}
``Microtensiometers'' are commonly used to measure the dynamic interfacial stress of microscopic spherical drops with radii of curvature on the order of $10 - 100~\upmu\text{m}$ \cite{kotula2015regular, javadi2012fast, alvarez2010microtensiometer, kazakov2008dilatational, manga2016measurements, alvarez2012interfacial, wilde2004proteins}. Being able to measure the dynamic interfacial stress of micron-scale systems provides a great advantage to the study of foams and emulsions, since the characteristic sizes of these systems typically range on the order of $10 - 100~\upmu\text{m}$ \cite{rodriguez2020asphaltene}. 

In these devices, a small spherical drop or bubble with $\text{Bo}\ll0.01$ is created at the tip of a capillary, which is itself connected to a pressure transducer \cite{kotula2015regular, javadi2012fast, alvarez2010microtensiometer}. Fluid is delivered to the tip of the capillary to create a drop/bubble, which is imaged using a magnified objective connected to either a regular or a high speed camera. The volume of the drop or bubble is controlled either directly, using a syringe pump, or indirectly, by adjusting the internal pressure of the drop/bubble via an external pressure head \cite{javadi2012fast, alvarez2012interfacial, kotula2015regular, russev2008instrument, liggieri2002measurement}. 

The dynamic surface tension for a simple interface $\sigma_{\alpha\beta}(\Gamma(t))$ can be determined from direct measurements of the pressure jump across the interface and the radius of the drop/bubble \cite{kotula2015regular, javadi2012fast, alvarez2010microtensiometer}. Since $\text{Bo}\ll0.01$, the hydrostatic pressure contribution can be neglected and Eq. \ref{eq.YoungLaplace} can be simplified to obtain the Young-Laplace equation for a spherical interface with a radius of curvature equal to $R_a$:
\begin{equation} \label{eq.YLsphere}
\Delta P_a = \frac{2 \sigma_{\alpha\beta}(\Gamma(t))}{R_a}.
\end{equation}
 
This same technique can also be used to determine the isotropic dynamic interfacial stress of a static, undeformed complex interface. 

Microtensiometry presents an advantage over other macroscopic techniques, which employ drops with radii of curvature on the order of millimeters, due to the faster adsorption times \cite{alvarez2010microtensiometer}. The use of micron-scale drops and bubbles not only requires smaller solution volumes than traditional methods, but also reduces the time required for an interface to reach its equilibrium configuration by almost an order of magnitude because the time scale for molecular diffusion is dependent on the radius of curvature of the interface, and is thus smaller for a convex curved interface compared to its planar counterpart \cite{alvarez2010microtensiometer, jaensson2018tensiometry}. 

As reported by Alvarez et al., the dynamic interfacial tension of large macromolecules (such as proteins and polymers) can be determined on the order of minutes or hours rather than days, as required with pendant drop tensiometry \cite{alvarez2010microtensiometer, kotula2015regular}. Furthermore, due to the smallness of the drops, high speed cameras with narrow fields of view can be used at frame rates upwards of 10,000 frames/s, which also allows this technique to be used to accurately study the adsorption dynamics of smaller molecules \cite{javadi2012fast}. 

 Microtensiometry also allows the user to determine whether the transport dynamics are governed by species diffusion or adsorption kinetics \cite{alvarez2010microtensiometer, kotula2015regular, ravera2010interfacial}. Surfactant transport to an initially clean interface is governed by three simultaneous transport processes: (1) diffusion of surfactant dissolved in the bulk  towards the fluid/fluid interface, (2) adsorption/desorption at the interface due to entropic effects, and (3) reorientation and reconfiguration of the adsorbed surfactant due to enthalpic effects \cite{alvarez2010microtensiometer, kotula2015regular}. Since diffusion is a function of the interfacial curvature, the dependence of the dynamic interfacial stress on the drop radius can be used to elucidate whether, at a particular bulk concentration and size, the transport dynamics diffusion limited \cite{alvarez2010microtensiometer}.

Microfluidic methods, which require the use of a convective bulk flow, have also been developed to measure interfacial stress at micron-scale interfaces. Additional details on this technique can be found in references \cite{narayan2018removing, chen2020size}.

\subsubsection{Dilatational rheology}
The dilatational rheology of complex fluid-fluid interfaces is correlated to the stability and lifetime of foams and emulsions \cite{kotula2015regular}. Understanding how complex fluid-fluid interfaces respond to area-changing deformations can also provide further insight to processes involving droplet break-up, nucleation, and coalescence. Dilatational deformations can be achieved using the drop/bubble setups outlined in the previous section. By changing the internal volume of a drop or bubble, the interface can be compressed or dilated either in a single step-wise manner or in a continuous oscillatory fashion. 

\begin{figure*}[]
\includegraphics[width=0.96\linewidth]{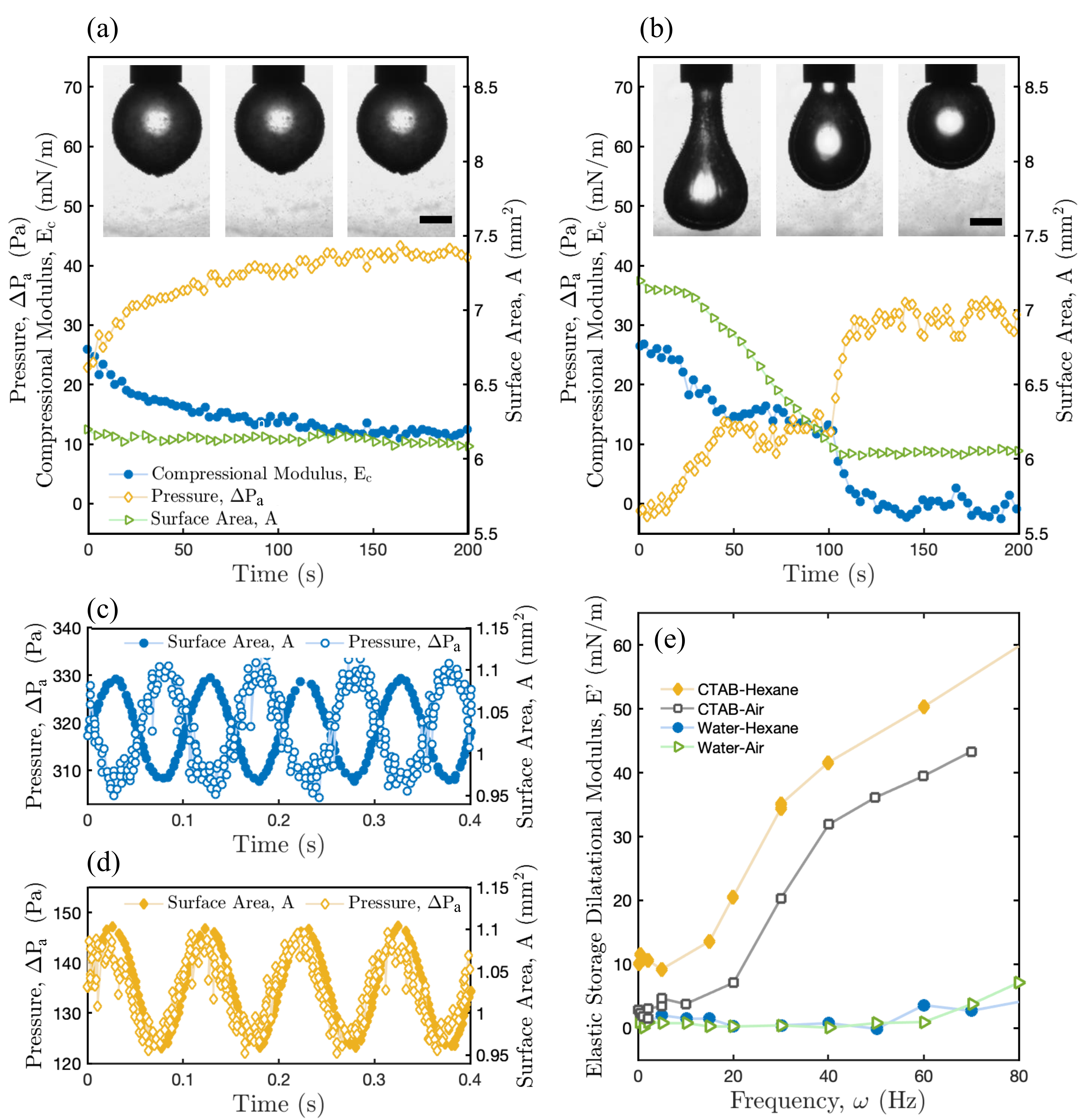} \caption{Top: Compressional step-strain relaxation profiles for a pendant water drop suspended in a $1~$mg/mL asphaltene in toluene solution ({\bf a.}) or a $1~$mg/mL asphaltene $+$ $2~$\% polymer in toluene solution ({\bf b.}). The drops are aged for 60 min and compressed at a flow rate of $0.1~\upmu\text{L}/\text{s}$. Snapshots are shown at 0, 100, and 200 s after stress relaxation begins (scale bar $=~0.5~$mm). Asphaltene-only interfaces show a time-dependent pressure relaxation but no shape change upon compression, whereas the polymer-laden system shows both a shape and a pressure relaxation. Results reproduced from Rodriguez-Hakim et al. \cite{rodriguez2020asphaltene}. Bottom: Surface area and Laplace pressure during small amplitude oscillations of a spherical drop of pure water ({\bf c.}) and a drop of aqueous $0.1~$mM CTAB solution ({\bf d.}) in hexane (frequency $=~10~$Hz). The in-phase component of the oscillations is used to calculate the storage elastic modulus, $E'$, for different water and CTAB interfaces. ({\bf e.}) Water-hexane interfaces are non-viscoelastic (phase shift$~=~\pi$ and $E'=0$) and CTAB-hexane interfaces exhibit a predominant elastic response (phase shift$~<~\pi/2$). Results reproduced from Javadi et al. \cite{javadi2012fast}.}
\label{fig:DilatationalRheology}
\end{figure*}

\subsubsection*{Step-strain}
During a step-strain measurement, a pendant bubble or drop is rapidly compressed/expanded by withdrawing/infusing fluid from it using a syringe pump \cite{nagel2017drop, danov2015capillary, carvajal2011mechanics}. This technique can be carried out with spherical or non-spherical geometries, and requires the use of the pendant bubble/drop setup for complex interfaces described in \bit{Section \ref{sec:IFT}}. Changes in the interfacial stress and drop geometry are measured as the drop/bubble is allowed to relax back to an equilibrium configuration \cite{nagel2017drop, danov2015capillary, carvajal2011mechanics}. Thus, step-strain experiments can be carried out using large areal strain deformations within the non-linear regime that improve the signal-to-noise ratio in the pressure transducer output and decrease the relative error in the drop area change calculations \cite{nagel2017drop}. 

Step-strain experiments consist of three steps: interface aging,  step-strain compression/expansion, and stress relaxation \cite{rodriguez2020asphaltene}. During the first step, a syringe pump is used to form a drop/bubble at the tip of a capillary, which is submerged in the bulk fluid. The initial drop/bubble shape can either be spherical  \cite{rodriguez2020asphaltene, kannan2018monoclonal, lin2016interfacial} or non-spherical \cite{nagel2017drop, danov2015capillary, pepicelli2019surface}. Spherical geometries are preferred when different compositions and interfacial tensions are being compared, as it is important to maintain a constant initial volume and surface area for all systems since experiments are often conducted within the nonlinear viscoelastic regime \cite{rodriguez2020asphaltene}. 

Once the drop is formed, the system is allowed to age for the desired aging time. After aging is complete, a step-strain compression/expansion is applied to the drop by using a syringe pump to withdraw/inject fluid until the final volume is reached. The applied flow rate can be thought of as analogous to a strain rate.  Thus, step-strain experiments can be conducted by varying the interface aging time and/or the compressional strain rate \cite{rodriguez2020asphaltene}. Once the drop reaches its final volume, it is allowed to relax back to an equilibrium shape and pressure. A time-dependent compressional relaxation modulus is calculated during this step \cite{rodriguez2020asphaltene, kannan2018monoclonal, lin2016interfacial}.  

Since step-strain dilatational rheology allows the use of non-spherical geometries, the measured interfacial stress can adopt a non-isotropic form as outlined in \bit{Section \ref{subsec:interfacialrheology}}. Depending on whether the measured stress is a scalar or a tensor, a time-dependent scalar or tensorial dilatational modulus can then calculated from the drop's surface area, radius, and the Laplace pressure jump across the apex, as follows \cite{rodriguez2020asphaltene, kannan2018monoclonal, lin2016interfacial},
\begin{equation} \label{eq.ModulusStepStrain}
    \boldsymbol{E}(t) = \frac{\Delta \boldsymbol{\sigma} (t) }{\Delta A / A_i} = \frac{\boldsymbol{\sigma}_i - \boldsymbol{\sigma}(t)}{(A_i - A(t))/A_i}.
\end{equation}
Here, $\boldsymbol{\sigma}_i$ and $A_i$ are the interfacial stress tensor and surface area before the step-strain deformation, and  $\boldsymbol{\sigma}(t)$ and $A(t)$ are the time-dependent values during relaxation.

Fig. \ref{fig:DilatationalRheology}(a-b) shows an example of different stress relaxation profiles that can be obtained with different complex interfaces, reproduced from Rodriguez-Hakim et al. \cite{rodriguez2020asphaltene}. In this example, the drop phase is composed of DI water and the bulk phase is a $1~$mg/mL asphaltene in toluene solution without (Fig. \ref{fig:DilatationalRheology}a) or with (Fig. \ref{fig:DilatationalRheology}b) the addition of a surface active co-polymer at $2~$wt\%. The results shown correspond to an interface aging time of $60~$min and a compressional flow rate of 0.1$~\upmu\text{L}/\text{s}$. The plots show the time evolution of the surface area $A$, apical Laplace pressure jump $\Delta P_a$, and apical compressional modulus $E_c$ during the relaxation step. Asphaltene-only interfaces show a time-dependent pressure relaxation but no shape change upon compression, whereas the polymer-laden system shows both a shape and a pressure relaxation. To simplify the analysis, the spatial dependence of the modulus was removed by only calculating the modulus value at the drop apex, where spherical symmetry holds and the elastic stresses are locally isotropic (i.e. $\sigma_{s} = \sigma_{\phi}$, as seen in Fig. \ref{fig:Sigmas_CMD}) \cite{rodriguez2020asphaltene, danov2015capillary, nagel2017drop}. 

Despite the differences in the temporal behavior of $E(t)$, two important parameters can be extracted from the curves: the initial compressional relaxation modulus $E_c(t=0)$ and the static (or equilibrium) compressional relaxation modulus, $E_c(t \rightarrow \infty)$ \cite{rodriguez2020asphaltene, kannan2018monoclonal, lin2016interfacial}. $E_c(t=0)$ represents the accumulation of elastic energy at the onset of compression \cite{rodriguez2020asphaltene, kannan2018monoclonal, lin2016interfacial} and  $E_c(t \rightarrow \infty)$ is the long-time equilibrium value of the surface elastic energy. Physically, it represents the degree of irreversibility of the film, or its solid-like character \cite{kannan2018monoclonal}. The lower $E_c(t \rightarrow \infty)$ is, the better the interface is at dissipating the accumulated elastic energy \cite{rodriguez2020asphaltene, kannan2018monoclonal, lin2016interfacial}. If $E_c(t \rightarrow \infty)$ is finite, as in Fig. \ref{fig:DilatationalRheology}a, the adsorbed species is irreversibly adsorbed onto the interface, forming a highly solid network that is associated with the long-term stability of emulsions. 

\subsubsection*{Oscillatory}
In oscillatory dilatational rheology, a sinusoidal change in the bubble/drop's surface area \cite{javadi2012fast} or pressure \cite{alvarez2012interfacial} is imposed via the injection and withdrawal of fluid. This technique requires the use of spherical drops or bubbles and is carried out using the microtensiometer setup outlined in \bit{Section \ref{sec:IFT}} or similar capillary pressure tensiometers \cite{alvarez2010microtensiometer, alvarez2012interfacial, javadi2012fast, russev2008instrument, liggieri2002measurement}. The drop/bubble is formed and remains undisturbed until adsorption equilibrium is established at the fluid/fluid interface \cite{javadi2012fast}. The interface is subjected to infinitesimal strain amplitudes, where the change in surface area, $\Delta A\lesssim10~$\% \cite{javadi2012fast}. This is particularly important for complex interfaces in order to ensure that a spherical geometry is maintained at all times. A pressure transducer is coupled to the capillary setup that allows for a simultaneous measurement of the surface area and internal pressure of the drop \cite{javadi2012fast, kotula2015regular, alvarez2010microtensiometer, ravera2010interfacial}. Measurement of the drop/bubble radius and internal pressure are sufficient to calculate the oscillatory dilatational moduli  \cite{javadi2012fast, kotula2015regular, alvarez2010microtensiometer, ravera2010interfacial}.

The dilatational modulus of an oscillating drop/bubble exhibits two contributions: an elastic part that represents the recoverable, or stored, energy of the interface (captured by the surface storage modulus, $E'$) and a viscous part that represents the dissipated energy (captured by the surface dilatational loss modulus, $E''$) \cite{freer2004relaxation, javadi2012fast, kotula2015regular, ravera2010interfacial}. $E'$ and $E''$ correspond to the real and imaginary parts of the complex surface dilatational modulus, $E^*$, where $E^* = E' + \text{i}E''$ \cite{freer2004relaxation, javadi2012fast, kotula2015regular, ravera2010interfacial}. The moduli values are functions of the oscillation frequency $\omega$; thus,
\begin{align}
E^*(\omega) &= \frac{\Delta \sigma}{\Delta A / A(t=0)}e^{i\Phi(\omega)}, \label{eq.ComplexMod} \\
E'(\omega) &= \frac{\Delta \sigma}{\Delta A / A(t=0)} \cos\Phi, \label{eq.StorageMod} \\
E''(\omega) &= \frac{\Delta \sigma}{\Delta A/ A(t=0)} \sin\Phi,  \label{eq.LossMod}
\end{align}
where $\Phi$ is the phase angle difference between the applied strain and the measured stress, $A(t=0)$ is the reference (initial) surface area, $\Delta A$ is the amplitude of the surface area strain, and $\Delta \sigma$ is the amplitude of the interfacial stress change \cite{freer2004relaxation, kotula2015regular, ravera2010interfacial}. Since the drops remain spherical at all times for small strain deformations, the interfacial stress remains isotropic even for complex interfaces (recall that for drops/bubbles with isotropic stress distributions, $\boldsymbol{\sigma}$ can be expressed as $\boldsymbol{\sigma} = \sigma\boldsymbol{I}_s$). Thus, for a constant radius of curvature, negligible gravitational effects (i.e $\text{Bo}\leq0.01$), and a spatially constant interfacial stress, the expression for $\sigma$ is given by Eq. \ref{eq.YLsphere}, where $\sigma = P_aR_a/2$.

Fig. \ref{fig:DilatationalRheology}(c-e) reproduces results obtained by Javadi et al. for simple and complex interfaces \cite{javadi2012fast}. Parts (c-d) of the figure show plots of the Laplace pressure jump $\Delta P_a$ and the surface area $A$ during oscillatory dilatational experiments with spherical drops composed of either pure water or a $0.1~$mM aqueous CTAB solution, respectively, in contact with a bulk hexane phase. This data can be used to compute the surface storage and loss moduli, $E'$ and $E''$, using Eqs. \ref{eq.YLsphere}, \ref{eq.StorageMod}, and \ref{eq.LossMod}. $E'$ is plotted in Fig. \ref{fig:DilatationalRheology}e for different drop and bulk compositions. The surface area and pressure oscillations in Fig. \ref{fig:DilatationalRheology}c for a pure water drop in hexane are completely out of phase (i.e a phase shift of $\pi$), since the water-hexane interface is simple and non-viscoelastic. As seen in Fig. \ref{fig:DilatationalRheology}e, simple interfaces such as water-hexane and water-air have moduli of zero. When CTAB adsorbs onto the water-hexane interface, it renders the interface viscoelastic (Fig. \ref{fig:DilatationalRheology}(d-e)). Since the oscillations in $A$ and $\Delta P_a$ are almost in phase, the interface has a predominant elastic character. Analogous results are seen for CTAB-air interfaces.

Due to geometric constraints, small amplitude oscillatory interfacial dilatational rheology is capable of determining the interfacial dilatational moduli for both simple and complex interfaces. It is also possible to conduct a frequency sweep by varying the oscillation frequency $\omega$ in order to see where the crossover between an elastic-dominated and a viscous-dominated response occurs \cite{freer2004relaxation, kotula2015regular}. 

This method has several limitations. Infinitesimal area strains are required, which may make it difficult to obtain accurate pressure readings from the transducer \cite{jaensson2018tensiometry}. Gas compressibility effects can also introduce spurious phase differences between the applied strain and the measured stress \cite{chandran2016impact}. Further, it is required that the pressure, stress, and area oscillations remain sinusoidal at all times, where the effect of higher order harmonics is mitigated \cite{jaensson2018tensiometry, kotula2015regular}. The magnitude of higher harmonics can be determined via a Fourier analysis of the oscillatory radius and pressure data \cite{loglio2005perturbation, kotula2015regular}. Kotula et al. specify an acceptable experimental criterion where the harmonic ratio (i.e the ratio of the second vs the first order harmonics) should be less than 0.1 \cite{kotula2015regular}.

In addition, the interfacial stress is computed using the Young-Laplace equation for a static interface, so it is assumed that the shape of the drop is in equilibrium at all times. This requires a slow, quasi-steady change in the drop shape, such that the capillary and Reynolds numbers are small \cite{jaensson2018tensiometry, kotula2015regular}. The capillary number, $\text{Ca}$, measures the relative contribution of viscous stresses arising from interfacial motion (where the drop/bubble apex translates a vertical distance $\Delta d$ during the period of an oscillation) versus dilatational stresses \cite{kotula2015regular}. The Reynolds number, $\text{Re}$, prescribes the relative importance between inertial and viscous stresses, where significant fluid inertia can cause additional pressure jumps across the interface \cite{kotula2015regular}. The operating dimensionless parameters for oscillatory interfacial dilatational rheology are summarized below, and a further discussion of the operating ranges can be found in Kotula et al. \cite{kotula2015regular}
\begin{align}
\text{Bo} &= \frac{\Delta\rho g R_a^2}{\sigma} \leq 10^{-2},\\
\text{Ca} &= \frac{\mu\omega\Delta d}{\sigma} \leq 10^{-6}, \\
\text{Re} &= \frac{\rho\omega R_a \Delta d}{\mu} \leq 10^{-1}.
\end{align}

\section{Foam stability}\label{sec:FoamStability}
\begin{figure*}[!th]
\includegraphics[width=\linewidth]{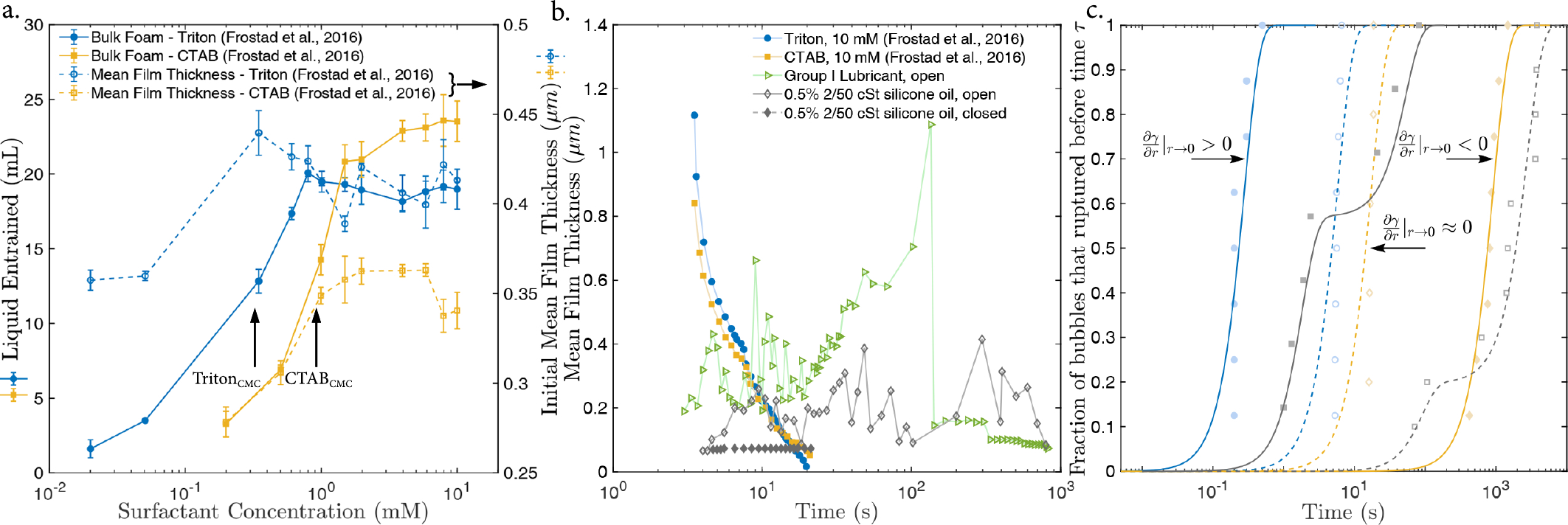} \caption{Insights into foam stability from single bubble measurements. ({\bf a.}) Demonstration of the positive correlation between bulk foam density and initial mean film thickness obtained from single bubble experiments as a function of surfactant concentration. Solid lines correspond to foam density measured from a foam rise test immediately after cutting of air injection, while the dashed lines correspond to the mean film thickness measured from single bubble experiments immediately after the bubble comes to rest at the air-liquid interface. Data obtained from Frostad et al. \cite{frostad2016dynamic}.  ({\bf b.}) Evolution of the mean film thickness, $\bar{h} = (\pi R_0^2)^{-1}\iint h(r,\theta) r dr d\theta$ (see Fig. \ref{fig:ExperimentalSetup} for definition of the variables).  The Triton and CTAB data are reproduced from Frostad et al. \cite{frostad2016dynamic}, while the Group I lubricant and silicone oil mixture data are reproduced from Suja et al. \cite{suja2018evaporation}. ({\bf c.}) Coalescence time distributions showing bubble stability in different systems. \protect\includegraphics[height=0.22 cm]{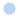} 10\% Toluene in $50$ cSt silicone oil (open), \protect\includegraphics[height=0.22 cm]{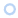} 10\% Toluene in $50$ cSt silicone oil (closed), \protect\includegraphics[height=0.22 cm]{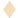} 0.5\% $2$ cSt in $50$ cSt silicone oil (open), \protect\includegraphics[height=0.22 cm]{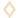} 0.5\% $2$ cSt in $50$ cSt silicone oil (closed), \protect\includegraphics[height=0.22 cm]{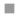} $10\;\mu m$ filtered lubricant, \protect\includegraphics[height=0.22 cm]{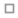} $1\;\mu m$ filtered lubricant. The silicone data is reproduced from Suja et al.\ \cite{suja2020symmetry}, while the lubricant data is reproduced from Suja et al.\ \cite{suja2020foam}. The silicone data shows the influence of the radial direction of Marangoni stresses on bubble stability, while the lubricant data shows the effect of the pore size of filters on bubble stability in filtered lubricants with antifoams. }\label{fig:FoamStability}
\end{figure*}

In this section we will discuss the recent developments in bubble and foam stability science facilitated by single bubble methods. 

\subsection{Foam Density}
Foam density, also referred to as the liquid fraction \cite{grassia2002foam, garrett2016science}, foam wetness \cite{stevenson2006wetness}, or quality, \cite{de1990rheology} is a measure of the amount of liquid entrained in the foam \cite{frostad2016dynamic}. Foam density is an important characteristic that has consequences for many industries such as food \cite{evans2002don}, froth flotation and extraction \cite{de1990rheology,simjoo2013foam}, and the lubricant industry \cite{binks2010non}. Traditionally, mechanistic studies on foam density are usually performed using bulk foam tests such as the foam rise test \cite{binks2010non,simjoo2013foam}. Recently, Frostad et al. \cite{frostad2016dynamic} have shown that single bubble experiments are a convenient platform to obtain mechanistic insights into foam density by establishing a correlation between the mean film thickness measured from single bubble experiments and the foam density measured from bulk foam experiments (Fig. \ref{fig:FoamStability}a). Subsequently, the same technique has been used by a number of researchers to probe the effects of interfacial properties on foam density \cite{kannan2019linking,lin2018influence}.

We will discuss two notable developments. Firstly, experiments by Frostad et al. \cite{frostad2016dynamic} have revealed the nuances in the role of Marangoni stresses in controlling foam density across different types of surfactants. As seen in (Fig. \ref{fig:FoamStability}a), at about $2~$mM concentration, the foam density in solutions with the surfactant CTAB crosses over the foam density of solutions with the surfactant Triton. This is very surprising as the surface tension of Triton is always lower than CTAB at a similar concentration.  A closer look at the evolution of the mean film thickness measured over single bubbles shows that despite trapping a thicker film as expected, bubbles in Triton solutions drain faster than in SDS solutions (Fig.\ref{fig:FoamStability} b). Even though the precise reason for this behavior is unknown, the relatively enhanced terminal drainage of Triton explains its lower foam density despite being capable of trapping a thicker film by generating larger Marangoni stresses.  
Secondly, experiments by Lin et al. \cite{lin2018influence} and Kannan et al. \cite{kannan2019linking} have presented a better understanding of the role of interfacial shear elasticity on the entrained film volume.  There exist contradicting conclusions on the effects of interfacial elasticity, with some studies correlating higher interfacial shear elasticity with higher entrained film volume while others finding no such correlation \cite{lin2018influence}. A resolution to these contradictions was presented by Kannan et al. \cite{kannan2019linking} by arguing that film drainage rates saturate above some critical value of the elastic modulus and that differences in film drainage can be perceived at lower values of the elastic modulus. Similar effects were reported for the film drainage as a function of interfacial viscosity of Newtonian interfaces. For films draining over solid domes, Bhamla et al. \cite{bhamla2014influence} observed an almost $100\%$ reduction in the drainage rate for a $10$-fold increase in the Boussinesq number from a value of $1$ (non-dimensional number proportional to the interfacial viscosity), while no significant changes in drainage were observed for a further increase in the Boussinesq number beyond a value of $10$. Both the above observations are most likely a result of the interface behaving as a no-slip surface above sufficiently high values of the interfacial modulus.

\subsection{Coalescence time distributions}\label{subsec:foamcoalescencetime}
As single bubble coalescence times are inherently stochastic, quantifying and comparing bubble coalescence times to bulk foam stability requires the use of appropriate statistical tools \cite{tobin2011public,lhuissier2012bursting,zheng1983laboratory}. One such tool is the coalescence time distribution (see Fig. \ref{fig:FoamStability}c and \bit{Section \ref{subsec:coalescencedistributions}}). Notable developments in this area are mentioned below. 

Firstly, recent results have shown that coalescence time distributions can be conveniently used to rank non-aqueous foam stability \cite{suja2018evaporation}. This is accomplished by constructing a series of distributions (eg. see coalescence times fit to Rayleigh distributions for silicone oil mixtures in Fig.\ref{fig:FoamStability}c), and inferring the relative position of the distributions. The farther right the distribution falls along the time axis, the more stable are the bubbles and consequently the more stable is the foam. The rationale for the varying foam stability in the silicone mixtures observed in Fig.\ref{fig:FoamStability}c is discussed in \bit{Section \ref{subsec:FoamStabilizationMechanisms}}. 

Secondly, coalescence time distributions have been shown to be sensitive to the presence of antifoams \cite{suja2020foam}. Coalescence times of naturally rupturing (without antifoams) bubbles are known to described by a single Weibull type distribution. However, in the presence of antifoams, we observe that bubble coalescence times are better described by mixture distributions (Fig.\ref{fig:FoamStability}c). This is not surprising as the coalescence time of a bubble is dependent on whether a bubble encounters an antifoam or not, with bubbles rupturing relatively quickly when antifoams are present. As a result, the measured coalescence times can fall under two different distributions with different means depending on the presence of antifoams. Further, the size of the antifoams also influence the coalescence time, with larger antifoams lowering the coalescence time. Both these effects can be seen in the coalescence time distributions (fit to Rayleigh distributions) of bubbles in antifoam-laden lubricants filtered using a $1~\upmu\text{m}$ and $10~\upmu\text{m}$ filter (Fig.\ref{fig:FoamStability}c). In the $1~\upmu\text{m}$ filtered lubricant, as a result of the very small filter pore size, the majority of the antifoam particles have been filtered out. Consequently a significant portion of the bubbles (those above the shoulder) never encounter an antifoam and remain stable for a longer time.  On the other hand, in the $10~\upmu\text{m}$ filtered lubricant, all bubbles encounter antifoams. However, due to a distribution of antifoam sizes in the lubricant, we again observe a mixture distribution. The means of the two distributions are most likely set by the two dominant antifoam sizes in the lubricant, with the distribution above the shoulder corresponding to bubbles ruptured by the smaller antifoam. Currently, efforts are underway to correlate the scale parameters and mixture ratios of the underlying Rayleigh distributions to the dominant antifoam sizes and their number densities \cite{suja2020foam}. 

As a concluding note, we highlight that the results presented in Fig. \ref{fig:FoamStability}c are for liquid antifoam droplets obeying the so called Garett's hypothesis \cite{garrett2016science,garrett1979effect}.  It would be worthwhile for future studies to investigate antifoams that do not adhere to the Garett's hypothesis and establish their influence on the coalescence time distributions.  

\subsection{Stabilization Mechanisms}\label{subsec:FoamStabilizationMechanisms}
 Single bubble experiments have played an important role in uncovering and mechanistically understanding foam stabilization mechanisms. A number of prior reviews have summarized the effects of interfacial rheology and traditional surfactant mediated Marangoni stresses in stabilizing bubbles \cite{langevin2015bubble, langevin2000influence, fameau2017non}. In this section we will focus only on the previously unreported mechanisms. Notable examples are presented below. 

Bubble stabilization by evaporation induced Marangoni flows has been the subject of a number of recent studies. Evaporation can drive Marangoni flows through changes in temperature as well as through changes in species concentration. The former, commonly referred to as thermocapillary Marangoni flows, is known to dictate bubble stability in highly volatile liquids with low specific heats \cite{menesses2019surfactant}. The later, commonly referred to as solutocapillary Marangoni flows \cite{rodriguez2019evaporation}, is known to alter the stability of bubbles in liquid mixtures with at least one volatile component. Evaporation driven solutocapillary Marangoni flows are known to increase bubble lifetimes in alcohol-water mixtures \cite{tuinier1996transient, lorenceau2020lifetime}. Interestingly, recent studies have revealed the important effect of solutocapillary flows on the stability of bubbles in non-aqueous systems such as lubricants \cite{suja2018evaporation, suja2020symmetry}. 

As shown in Fig.\ \ref{fig:FoamStability}c using mixtures of silicone oils, bubble stability depends on the radial direction of the Marangoni stresses induced by evaporation. Bubbles are stabilized when the Marangoni stresses compete against capillary flows and drive fluid to the bubble apex, while bubbles are destabilized when Marangoni stresses drive fluid away from the bubble apex. An interesting signature of the former case is the spontaneous cyclic dimple formation and dissipation resulting from the competition of Marangoni flows that drive fluid to the apex of the bubble and capillary flows that thin down the film \cite{shi2020oscillatory, velev1993spontaneous, suja2018evaporation}. As a result, dramatic fluctuations are observed in the film thickness of bubbles, along with a marked increase in their life time (see the data for a Group I lubricant and a silicone oil mixture in Fig.\ref{fig:FoamStability}b). When evaporation is suppressed, for instance by sealing the system, capillary forces steadily drain the film without competition and no fluctuations are observed. As expected, for closed systems, bubble stability decreases if evaporation is stabilizing and vice versa (Fig.\ \ref{fig:FoamStability}c).

\subsection{Bubble rupture dynamics}
Single bubble experiments have also played a pivotal role in establishing the rupture dynamics of bubbles. Good discussions on hole opening kinetics \cite{muller2007experimental, pandit1990hydrodynamics, muller2009comparison}, topological changes \cite{robinson2001observations, muller2007experimental, muller2009comparison}, and fragmentation dynamics \cite{lhuissier2012bursting,pandit1990hydrodynamics} are available in the literature. Here we will briefly comment on the recent developments in this area. 

Firstly, recent studies have revealed the influence of bulk elasticity in the hole opening kinetics of bubbles in a number of systems such as Boger fluids \cite{tammaro2018elasticity}, wormlike micelles \cite{sabadini2014elasticity}, and polymer melts \cite{debregeas1998life}. In all cases, at short times, bulk elasticity was revealed to increase the hole opening velocity by as much as $10^4$ times as compared to a Newtonian fluid of similar viscosity. This is expected as the elastic stresses that build up during bubble formation aid the capillary stresses in rupturing the bubble, leading to an increase in the rupture velocity. 

Secondly, recent research has also improved our understanding of the topology of bubbles during rupture. Notably, Debr\'egeas et al. \cite{debregeas1998life} have shown that during rupture, buckling instabilities can occur on the surface of bubbles in polymer melts. Sabadini et al., \cite{sabadini2014elasticity} on the other hand, have interestingly reported a complete absence of a rim (the tip of the expanding hole where liquid accumulates) in bubbles rupturing in viscoelastic wormlike micellar solutions. The reason for this is currently unknown. 

Thirdly, a number of studies have focused on fragmentation dynamics \cite{lhuissier2012bursting,poulain2018biosurfactants}.  The retracting fluid at the rim is known to fragment via a sequence of hydrodynamic instabilities, namely a Rayleigh-Taylor instability generating the ligaments at the bubble rim followed by a Rayleigh-Plateau instability generating droplets from the ligaments. For bubbles in simple liquids, the mean size of these generated droplets $\langle d \rangle$  was shown by Lhuissier and Villermaux to scale with the mean thickness of the film $\bar{h}$ (see Fig.\ref{fig:FoamStability} caption for the mathematical definition of $\bar{h}$) as $\langle d \rangle \sim \bar{h}^{5/8}$, and from mass conservation, the number of drops $N$ to scale as $N \sim \bar{h}^{-7/8}$. Building on this result, Poulain et al. \cite{poulain2018biosurfactants} have shown that bacterial secretions reduce the size and increase the number of droplets released during bubble rupture by lowering the film thickness at rupture. As a result, these pathogens spread more readily by taking advantage of the mechanics of bubble rupture. In the future, it would be worthwhile for studies to further investigate the impact of interfacial properties, especially the effects of interfacial rheology, on the dynamics of bubble rupture \cite{bico2015cracks}.

\begin{figure*}
\centering
\includegraphics[width=\linewidth]{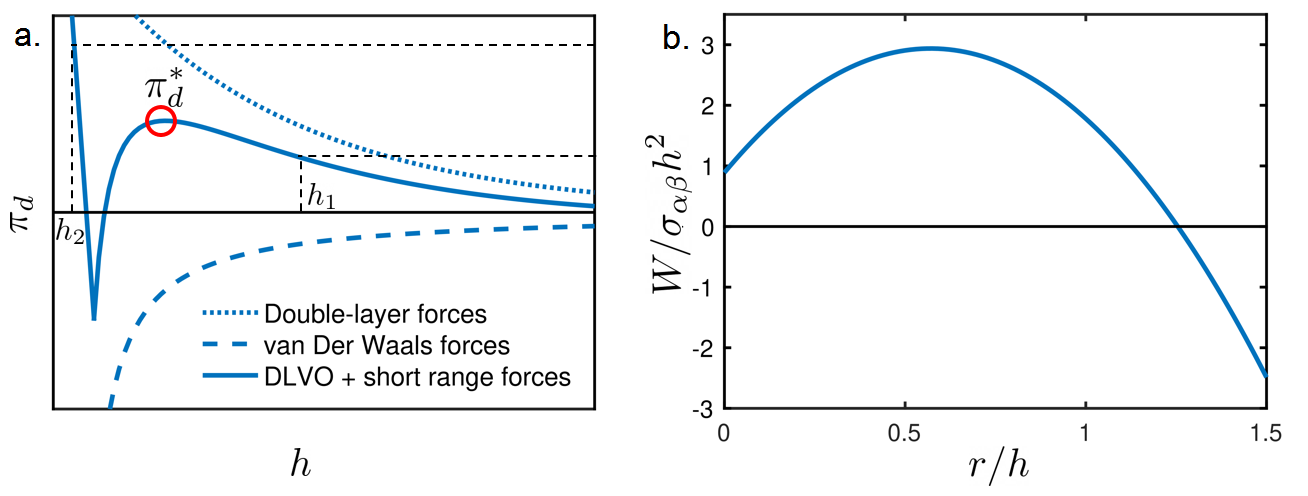} \caption{({\bf a.}) Disjoining pressure versus film thickness $h$, where the dotted curve represent the double-layer forces (the electrostatic -repulsive- contribution) and the dashed curve represents the van Der Waals forces (attractive). The solid curve represents the addition of these two contributions (DLVO forces) plus the short range forces. If the pressure in the thin film is smaller than the local maximum, $\pi_d^*$, DLVO fores dominate and the equilibrium thickness $h$ is of the order of hundreds of nm ($h_1$). If the pressure is bigger than $\pi_d^*$, $h$ is much smaller (of the order of a few nm, $h_2$), short range forces dominate and, essentially, the two interfaces form a bilayer. ({\bf b.}) Non-dimensional free energy of the nucleated hole as a function of its non-dimensional size, calculated from the de Vries theory \cite{deVries1958foam}. The hole nucleation is energetically unfavourable until a critical $r/h$ value is reached (about 0.5). Here $r$ is the radius of the hole, $h$ is the film thickness, $W$ is the free energy, and $\sigma_{\alpha\beta}$ is the interfacial tension. The hole spontaneously increases in radius once that critical value is exceeded.}\label{jt_DLVO}
\end{figure*}

\section{Emulsion stability}\label{sec:EmulsionStability}
The physical mechanisms governing the stability of an emulsion are not yet fully understood, but there exist a number of theories confirmed by experiments that have shed light on this problem for over more than a century. In this section we present some of the most relevant and established models dealing with emulsion stability and coalescence, as well as more recent advances and potential developments of the single drop techniques.

\subsection{Stabilization mechanisms and film rupture}\label{subsec:stableandunstable}
A stable emulsion can be formed under some conditions, and by means of different physical mechanisms. The most common procedure to increase the stability of emulsions is the addition of surface active species in sufficient quantity to form dense surface layers \cite{Langevin2019coalescence}. Stable thin films of constant thickness can then be formed, preventing adjacent droplets from coalescing. In particular, it was proposed \cite{Hildebrand1941emulsion} and experimentally demonstrated \cite{Traykov1977hydrodynamics} that the added surfactant must be soluble in the continuous phase and insoluble in the disperse phase in order to optimize the increase in stability. The relevant stabilization mechanism is the well known Marangoni flow: when two droplets approach and come into contact (or a droplet and a planar interface), surfactant molecules are driven towards the film perimeter, creating gradients in surface concentration, and in turn, surface tension gradients. When the surfactant is soluble in the disperse phase, there is a source of surfactant molecules to rapidly replenish the surface, eliminating the surface tension gradients. Hence, Marangoni flows are suppressed and the film thins faster. On the contrary, when the surfactant is soluble only in the continuous phase and the film is thin enough, there are not enough surfactant molecules available to replenish the surface. Hence, Marangoni flows that oppose the film thinning are sustained, and the film thins at a slower rate \cite{Langevin2019coalescence}.

In general, ionic surfactants are more efficient in stabilizing emulsions. An intuitive explanation of this observation is given in Fig.\ \ref{jt_DLVO}a, where we represent the disjoining pressure, $\pi_d$, versus the film thickness when a monolayer of an ionic surfactant is present in both interfaces \cite{Langevin2019coalescence,Ivanov1997stability,karakashev2008effect}. At large film thicknesses, the interaction between the interfaces is governed by the addition of electrostatic repulsion (screened-Coulomb or Yukawa potential) and van der Waals attraction, known as DLVO (Derjaguin-Landau-Verwey-Overbeek) forces. At much smaller film thicknesses, short range forces govern the dynamics. If the pressure in the thin film is smaller than the local maximum of the disjoining pressure represented in Fig. \ref{jt_DLVO}a by a red circle ($\pi_d^*$), the film would thin until it reaches an equilibrium value of the order of hundreds of nanometers. If the pressure is higher, the equilibrium thickness is much smaller, where the fluid separating the monolayers has been fully removed and a bilayer is formed.

It is well known that solid particles located on the liquid/liquid interface can increase emulsion stability, forming the so-called Pickering emulsions \cite{pickering1907emulsions}. There exists strong evidence that the physical mechanism arresting coalescence in Pickering emulsions is the formation of a steric barrier by the particles \cite{monegier2015influence,menon1988characterization,binks2002solid,aveyard2003emulsions,yang2017overview}. This mechanism requires the adsorption of the particles at the interface, which is possible only when the three phase contact angle is close to $90^\circ$. Hence, the amphiphilic character of the particles facilitates the stabilization of the emulsion. The main application of Pickering emulsions, extensively used in the last decades, is the fabrication of nanomaterials such as microspheres and microcapsules, with direct applications in the food or pharmaceutical industries \cite{chen2010novel,williams2015inorganic,frelichowska2009pickering}. For a review on Pickering emulsions focused on the different types of emulsifying particles  and the nanomaterials fabricated from Pickering emulsions, the reader is addressed to Yang et al. \cite{yang2017overview}.

In spite of the fact that the coalescence process is not fully understood, the physical mechanisms leading to the apparition and eventual nucleation of a hole in the thin film have been examined for decades. De Vries \cite{deVries1958foam} studied the energetics of hole nucleation, finding that there exists a critical hole size below which hole growth is energetically unfavorable. This theory is based on the calculation of the increment in surface area associated with the hole growth, where there is both a loss in surface area given by $2\pi r^2$ (where $r$ is the radius of the hole) and an increase in surface area due to the formation of a hole rim. Assuming that the hole rim is perfectly circular and the surface tension is uniform, the resulting non-dimensional free energy of the nucleated hole as a function of its non-dimensional size is represented in Fig. \ref{jt_DLVO}b, where the free energy, $W$, has been made dimensionless by $\sigma_{\alpha\beta} h^2$, $\sigma_{\alpha\beta}$ being the interfacial tension and $h$ the thin film thickness.

\begin{figure*}
\centering
\includegraphics[width=\linewidth]{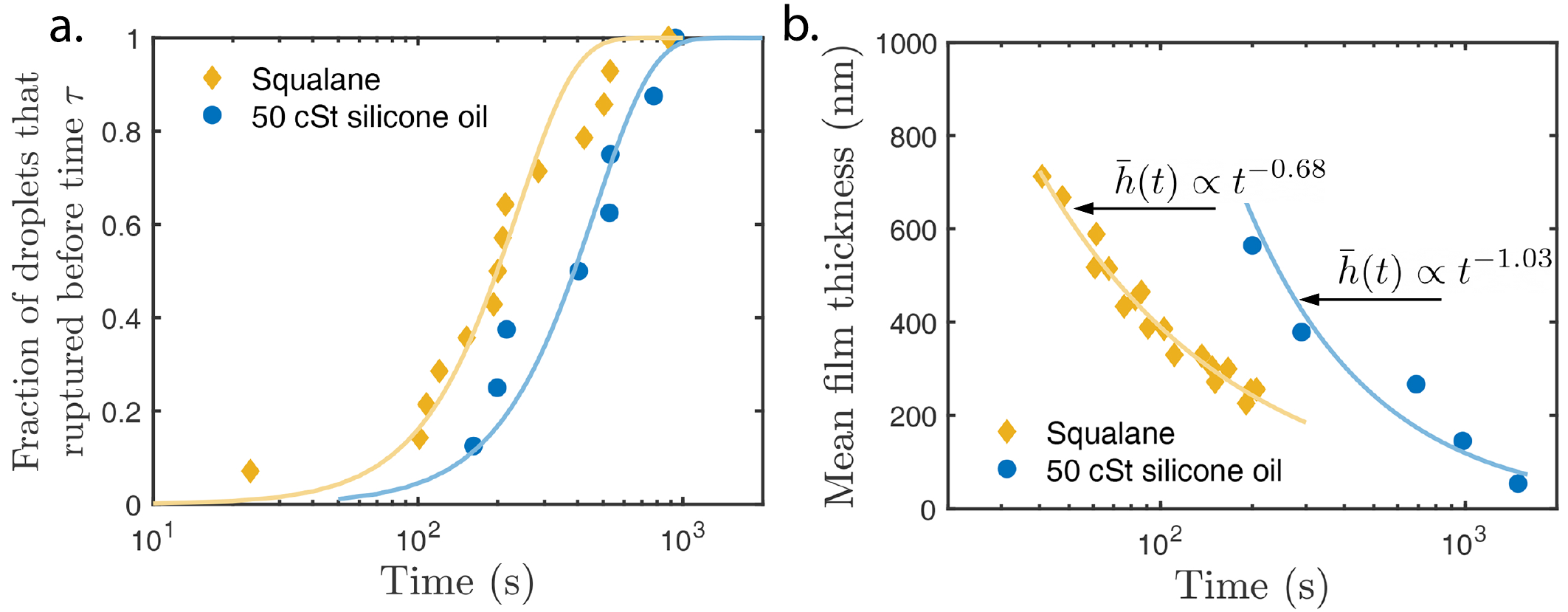} \caption{({\bf a.}) Coalescence time distributions of oil droplets against a flat oil/water interface. The symbols represent the experimental data, and the curves show the corresponding Rayleigh distributions. ({\bf b.}) Mean film thickness ($\bar{h}$) versus time for the same two samples. The curves show the best fit to a function of the form $\bar{h}(t)\propto t^{-\xi}$ where the the best fit values for $-\xi$ are -1.03 and -0.68 for the silicone oil and the squalane, respectively.}\label{jt_results}
\end{figure*}

Other models have been proposed \cite{Kashchiev1980nucleation}, accounting, for instance, for the spontaneous curvature of the surfactant molecules \cite{Kabalnov1996macroemulsion,Kabalnov1998coalescence}. However, all these models predict an activation energy orders of magnitude higher than $k_BT$ for the typical values of the equilibrium thickness ($\sim100\mbox{\,nm}$). Therefore, a reasonable coalescence mechanism for emulsions stabilized by ionic surfactants is as follows \cite{Langevin2019coalescence}: the local maximum in Fig.\ref{jt_DLVO}a arising from the electrostatic potential is decreased due to surface concentration fluctuations, allowing the film to locally thin to a much smaller thickness, where short range forces dominate the dynamics. The activation energy is not much larger than $k_BT$ now, such that hole nucleation is possible. Following the nomenclature by de Gennes \cite{deGennes2001some}, this mechanism represents the classical view of \textit{intrinsic} coalescence. However, in many applications, the most favorable coalescence mechanism is different in nature: analogous to the effect of antifoams discussed in \bit{Section \ref{subsec:foamcoalescencetime}}, a particle located in the thin film separating two droplets can bridge the film and facilitate coalescence, producing the so-called \textit{extrinsic} coalescence.

\subsection{Single drop experimental approach}
As explained in \bit{Section \ref{sec:Methods}}, the experimental setup for single drop and single bubble experiments is essentially identical. Moreover, the main observables are coalescence times and interferometric images in both cases. Since the experimental details and the fundamental details of coalescence time distributions and film thickness reconstruction have already been discussed in the preceding sections, we will confine ourselves to present some recent results to highlight the applicability of the technique to the emulsion problem. At the end of this section, we will examine the film rupture dynamics in the case of emulsions, discussing the potential application of single drop techniques to this question.

\subsubsection{Coalescence time distribution and drainage rate}
Fig.\ref{jt_results}a shows coalescence time distributions for squalane/water and lubricant oil/water systems. The experimental data follow a distribution similar to that discussed in the \bit{Section \ref{subsec:coalescencedistributions}}, where a Rayleigh distribution reasonably fits the data. The technique is also suitable to rank emulsion stability, as can be inferred from the two distinct distributions observed; in other words, coalescence is a random process and the experimental data show a remarkable dispersion, but a significant number of experiments allows one to estimate the probability of coalescence, which is an intrinsic property of the system.

As is the case with foams, the interferometric images enable the thin film thickness reconstruction. At the cost of a relatively manual process, it is possible to calculate the topography of the film and its evolution with time. Fig. \ref{jt_results}b shows the mean film thickness versus time for the same two systems as in Fig. \ref{jt_results}a. In both cases, $t=0$ is given by the instant in which the stage controlling the drop vertical position stops moving, and the maximum thickness that can be measured is optically limited ($\sim1\,\upmu\mbox{m}$). It is worth mentioning that, in these experiments, no dimple was observed, which does not necessarily mean that a dimple was not formed during the initial stages where $h>1\,\upmu\mbox{m}$. As can be seen in Fig. \ref{jt_results}b, the mean film thickness reasonably follows a power law of the form $\bar{h}(t)\propto t^{-\xi}$, where the best fit values for $-\xi$ (-1.03 and -0.68 for the silicone oil and the squalane, respectively) interestingly coincide with those expected respectively for a plug flow and poiseuille flow inside the thin film \cite{Frostad2013scaling}.


\begin{figure*}
\centering
\includegraphics[width=\linewidth]{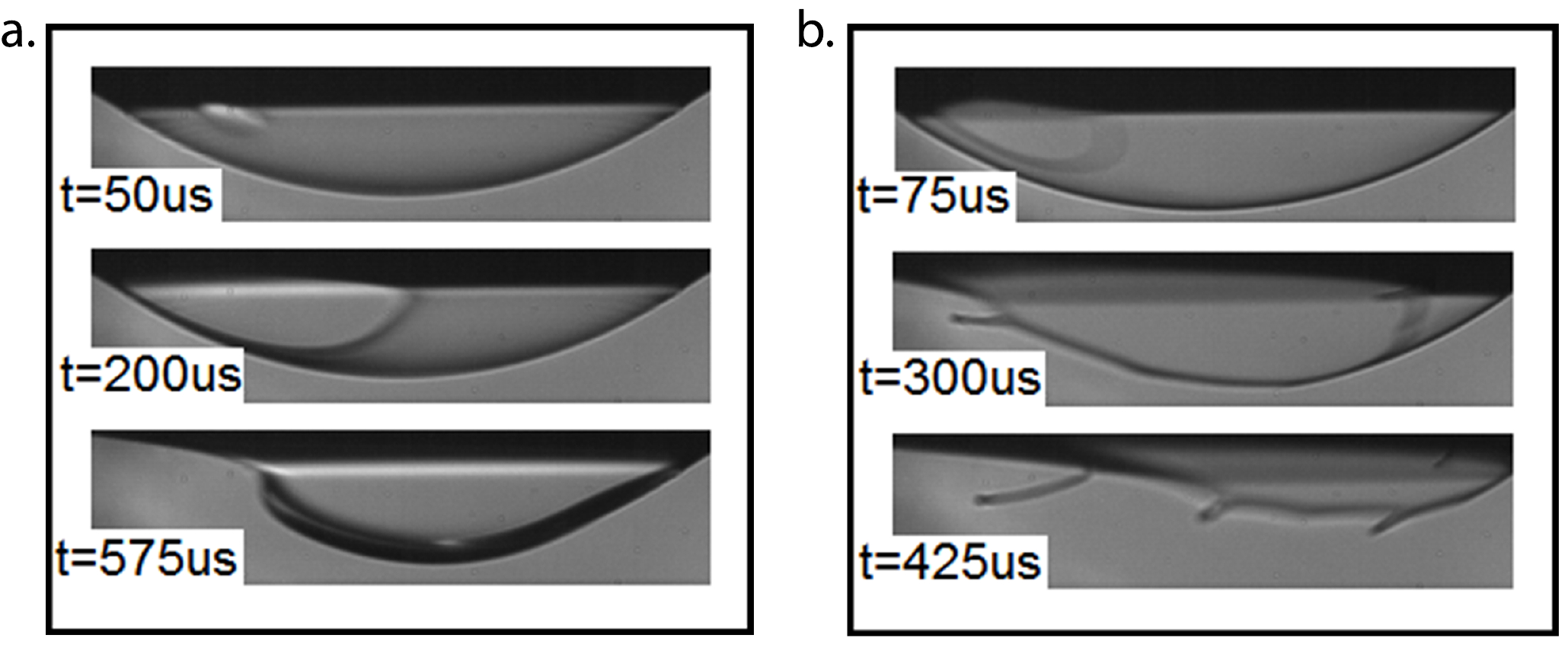} \caption{Snapshots of the film rupture acquired by means of a high speed camera used as the side camera in Fig. \ref{fig:ExperimentalSetup}. In both cases, the lower and upper phases are DI water, and the thin film is a 50\,cSt pure silicone oil. Two distinct regimes are observed. In the left images ({\bf a.}), the film ruptured at a much larger thickness than that corresponding to the right images ({\bf b.}), so that $Oh_{\text{right}}\gg Oh_{\text{left}}$. A rim is clearly observed in the left images, growing in size as the film retracts and fluid volume accumulates. On the contrary, tendrils of oil are left behind on the hole perimeter in the right images, where the $Oh$ number is lower.}\label{jt_highspeed}
\end{figure*}

\subsubsection{Drop rupture dynamics and retraction speed}
The experimental setup represented in Fig. \ref{fig:ExperimentalSetup}c is also an excellent platform to study the hole nucleation mentioned in \bit{Section \ref{subsec:stableandunstable}}. The substitution of the side camera by a high speed camera allows one to track the growth of the hole, as shown in Fig. \ref{jt_highspeed}. The retraction of the thin film during coalescence is a rich phenomenon that has been studied over decades \cite{Kavehpour2015coalescence}, with an increased interest after the development of digital high speed cameras opened the door towards its direct visualization. There exist different regimes, where the Ohnesorge number, defined as,
\begin{align}
Oh=\frac{\mu}{\sqrt{\rho\sigma_{\alpha\beta} h_r}},
\end{align}
tells us whether the system pertains to the inertial regime ($Oh\ll 1$) or the viscous regime ($Oh>1$) \cite{Kavehpour2015coalescence}. Here $\mu$ is the dynamic viscosity of the continuous phase,  $\rho$ is the density of the continuous phase, $\sigma_{\alpha\beta}$ is the surface tension, and $h_r$ is the film thickness at rupture. 

In the examination of the inertial regime, several research efforts have found an increase of the radius of the hole following $r\sim \sqrt{t}$ \cite{Fullana1999stability,Menchaca2001coalescence,Duchemin2002inviscid,Aarts2005hydrodynamics,Thoroddsen2005coalescence,Aryafar2006inertia}, with the mass of the retracting film accumulating in a torus-shaped rim. For $Oh\gtrsim 1$, a model was proposed where the radius of the hole follows $r/R\sim t/t_v(\ln(t/t_v))$, being $R$ the drop radius and $t_v$ the viscous time defined as \cite{eggers1999coalescence},
\begin{align}
    t_v=\frac{R\mu}{\sigma_{\alpha\beta}}.
\end{align}
Later experiments found a relationship $r \sim t$ for this regime \cite{Menchaca2001coalescence,Aarts2005hydrodynamics,Thoroddsen2005coalescence} and, very recently, Zhang \textit{et al.} \cite{Zhang2019initial} have used a single drop setup to confirm the above-mentioned $r\sim\sqrt{t}$ and $r\sim t$ relationships for the inertial and viscous regimes, respectively. Aryafar and Kavehpour \cite{Aryafar2008hydrodynamic} pointed out a possible explanation for the observed discrepancy in the high $Oh$ regime: the rim of the hole becomes unstable, forming tendrils that eventually produce micron sized droplets. Since the hole radius is normally measured from side images (measuring the length of the neck between the two droplets or the droplet and the planar interface), this instability could not be observed in most of the experiments cited above.

Another instance in which the visualization of the neck (and not the hole) may not fully capture the complexity of the hole growth is the case of the two interfaces entrapping a thin film that has a different interfacial tension at either liquid-liquid interface due to, for example, a different aging process. Malmazet \textit{et al.} \cite{deMalmazet2015coalescence} conducted coalescence experiments where a water droplet is released from a capillary immersed in oil. Once released, the droplet descends and reaches an oil/water interface, which has been aged and shows a lower interfacial tension. When coalescence happens, they observed the rim bending towards the inside of the drop, i.e., towards the interface showing a higher interfacial tension.

Images of a torus-shaped rim and an unstable rim with tendrils are shown in Fig. \ref{jt_highspeed}. Note that these images were obtained by means of the substitution of the side camera represented in Fig. \ref{fig:ExperimentalSetup}c by a high speed camera. In other words, the setup can still simultaneously work as an interferometer, providing the topography of the film when coalescence takes place. Since the film thickness is of critical importance to the mechanism leading to the observed instability at high $Oh$ numbers \cite{Kavehpour2015coalescence}, the single drop setup is a promising tool to further analyze this phenomenon.

\section{Conclusion and Outlook}
Single bubble/drop techniques have improved our understanding of foam and emulsion physics by providing insights that complement those obtained from the bulk foam/emulsion and single film tests. Typical bubble/drop setups contain an arrangement to form bubbles/drops of controlled volume (often supported on a capillary), cameras to visualize the shape of the bubble/drop, pressure transducers to monitor the internal pressure, and an arrangement (usually based on interferometry) to measure the spatiotemporal film thickness evolution between the interacting interfaces of bubbles or drops. Major measurables from single bubble/drop experiments include  coalescence times and their distributions, the spatiotemporal film profiles both during drainage (utilizing thin film interferometry) and rupture (utilizing a high speed imaging camera), and interfacial rheology.      

Single bubble/drop techniques will continue to be an important tool for studying foams and emulsions. Future work in this area can be split into two categories. Firstly, efforts can be aimed at improving the single bubble/drop setups and the associated protocols. These include (a) improving film thickness measurement tools (interferometry or otherwise) for studying emulsions with low refractive indices (eg. flurosilicone-water emulsions), (b) improving the robustness and spatiotemporal resolution of automated film thickness measurement tools (eg. for interferometry), (c) developing a generalized theory for describing coalescence time distributions across bubbles, anti-bubbles, and drops, and (d) investigating the role of bubble size (super or sub hemispherical cap) on the accuracy of dilatational rheology measurements.  Secondly, efforts can be aimed at utilizing single bubble/drop setups to resolve unanswered questions in foam and emulsion physics. These include (a) investigating antifoam mechanics in non-aqueous systems, (b) characterizing evaporation driven solutocapillary bubble destabilization, and (c) investigating the effects of interfacial rheology on the dynamics of bubble rupture.

\section*{Acknowledgements}
We thank John Frostad and Simone Bochner de Araujo for constructing the experimental setups reported in this manuscript, and Prem Sai for the schematic illustrations in the manuscript.  This study was partially supported by grants from Shell and the Beijing Welltrailing Science and Technology Company.

\appendix
\section{Supplementary Data}
The supplementary data can be obtained online at . The MATLAB\textsuperscript{\textregistered} codes for computing the coalescence time distributions can be obtained at 
\begin{enumerate}
    \item Coalescence time curves -\\
    \href{https://github.com/vcsuja/Coalescence-Time-Distributions}{https://github.com/vcsuja/Coalescence-Time-Distributions} 
    \item Color Map generator -\\ \href{https://github.com/vcsuja/ColorMapGenerator}{https://github.com/vcsuja/ColorMapGenerator} 
\end{enumerate}

\bibliographystyle{vancouver}
\bibliography{Reference}

\begin{thebibliography}{100}

\bibitem{frostad2016coalescence}
Frostad JM, Paul A, Leal LG.
\newblock Coalescence of droplets due to a constant force interaction in a
  quiescent viscous fluid.
\newblock Physical Review Fluids. 2016;1(3):033904.

\bibitem{pugh2016bubble}
Pugh RJ.
\newblock Bubble and foam chemistry.
\newblock Cambridge University Press; 2016.

\bibitem{garrett2016science}
Garrett PR.
\newblock The science of defoaming: theory, experiment and applications. vol.
  155.
\newblock CRC Press; 2016.

\bibitem{patino2008implications}
Patino JMR, Sanchez CC, Ni{\~n}o MRR.
\newblock Implications of interfacial characteristics of food foaming agents in
  foam formulations.
\newblock Advances in Colloid and Interface Science. 2008;140(2):95--113.

\bibitem{qian2009study}
Qian S, Wu Z, Zheng H, Geng Y.
\newblock Study on riboflavin recovery from wastewater by a batch foam
  separation process.
\newblock Separation Science and Technology. 2009;44(11):2681--2694.

\bibitem{evans2002don}
Evans DE, Sheehan MC.
\newblock Don't be fobbed off: The substance of beer foam—A review.
\newblock Journal of the American Society of Brewing Chemists.
  2002;60(2):47--57.

\bibitem{kitchener1984froth}
Kitchener J.
\newblock The froth flotation process: past, present and future-in brief.
\newblock In: The Scientific Basis of Flotation. Springer; 1984. p. 3--51.

\bibitem{suja2018evaporation}
Suja VC, Kar A, Cates W, Remmert S, Savage P, Fuller G.
\newblock Evaporation-induced foam stabilization in lubricating oils.
\newblock Proceedings of the National Academy of Sciences.
  2018;115(31):7919--7924.

\bibitem{suja2020foam}
Suja VC, Kar A, Cates W, Remmert S, Fuller G.
\newblock Foam stability in filtered lubricants containing antifoams.
\newblock Journal of Colloid and Interface Science. 2020;567:1--9.

\bibitem{sawicki1992high}
Sawicki GC.
\newblock High-Performance Antifoams for the Textile Dyeing Industry.
\newblock In: Surfactant Sci. Ser.. vol.~45. Marcel Dekker New York; 1992. p.
  193.

\bibitem{tsang2006novel}
Tsang YF, Chua H, Sin S, Tam C.
\newblock A novel technology for bulking control in biological wastewater
  treatment plant for pulp and paper making industry.
\newblock Biochemical Engineering Journal. 2006;32(3):127--134.

\bibitem{leal2007emulsion}
Leal-Calderon F, Schmitt V, Bibette J.
\newblock Emulsion science: basic principles.
\newblock Springer Science \& Business Media; 2007.

\bibitem{chappat1994some}
Chappat M.
\newblock Some applications of emulsions.
\newblock Colloids and Surfaces A: Physicochemical and Engineering Aspects.
  1994;91:57--77.

\bibitem{mandal2010characterization}
Mandal A, Samanta A, Bera A, Ojha K.
\newblock Characterization of oil- water emulsion and its use in enhanced oil
  recovery.
\newblock Industrial \& Engineering Chemistry Research.
  2010;49(24):12756--12761.

\bibitem{tadros2011colloids}
Tadros TF.
\newblock Colloids in paints. vol.~6.
\newblock John Wiley \& Sons; 2011.

\bibitem{harika2011impact}
Harika E, Helene M, Bouyer J, Fillon M.
\newblock Impact of lubricant contamination with water on hydrodynamic thrust
  bearing performance.
\newblock M{\'e}canique \& industries. 2011;12(5):353--359.

\bibitem{bochner2017droplet}
Bochner~de Araujo S, Merola M, Vlassopoulos D, Fuller GG.
\newblock Droplet coalescence and spontaneous emulsification in the presence of
  asphaltene adsorption.
\newblock Langmuir. 2017;33(40):10501--10510.

\bibitem{azim2011demulsifier}
Azim AAAA, Abdul-Raheim ARM, Kamel RK, Abdel-Raouf ME.
\newblock Demulsifier systems applied to breakdown petroleum sludge.
\newblock Journal of Petroleum Science and Engineering. 2011;78(2):364--370.

\bibitem{umar2018review}
Umar AA, Saaid IBM, Sulaimon AA, Pilus RBM.
\newblock A review of petroleum emulsions and recent progress on water-in-crude
  oil emulsions stabilized by natural surfactants and solids.
\newblock Journal of Petroleum Science and Engineering. 2018;165:673--690.

\bibitem{astm2013standard}
D892-13 A. Standard Test Method for Foaming Characteristics of Lubricating
  Oils.
\newblock ASTM International West Conshohocken, PA; 2013.

\bibitem{denkov2004mechanisms}
Denkov ND.
\newblock Mechanisms of foam destruction by oil-based antifoams.
\newblock Langmuir. 2004;20(22):9463--9505.

\bibitem{flenderfoamISO}
{ISO 12152:2012-08}.
\newblock { Lubricants, Industrial Oils and Related Products -- Determination
  of the Foaming and Air Release Properties of Industrial Gear Oils Using a
  Spur Gear Test Rig/ Flender Foam Test Procedure}.
\newblock Geneva, CH: International Organization for Standardization; 2012.

\bibitem{astm2019standard}
D1401-19 A. Standard Test Method for Water Separability of Petroleum Oils and
  Synthetic Fluids.
\newblock ASTM International West Conshohocken, PA; 2019.

\bibitem{kamkar2020polymeric}
Kamkar M, Bazazi P, Kannan A, Chandran~Suja V, Hejazi SH, Fuller GG, et~al.
\newblock Polymeric-Nanofluids Stabilized Emulsions: Interfacial versus Bulk
  Rheology.
\newblock Journal of Colloid and Interface Science. 2020;.

\bibitem{lobo2002coalescence}
Lobo L, Svereika A, Nair M.
\newblock Coalescence during emulsification: 1. Method development.
\newblock Journal of colloid and interface science. 2002;253(2):409--418.

\bibitem{joscelyne2000membrane}
Joscelyne SM, Tr{\"a}g{\aa}rdh G.
\newblock Membrane emulsification—a literature review.
\newblock Journal of Membrane Science. 2000;169(1):107--117.

\bibitem{pugh1996foaming}
Pugh R.
\newblock Foaming, foam films, antifoaming and defoaming.
\newblock Advances in Colloid and Interface Science. 1996;64:67--142.

\bibitem{friberg2010foams}
Friberg SE.
\newblock Foams from non-aqueous systems.
\newblock Current opinion in colloid \& interface science. 2010;15(5):359--364.

\bibitem{pugh2005experimental}
Pugh R.
\newblock Experimental techniques for studying the structure of foams and
  froths.
\newblock Advances in Colloid and Interface Science. 2005;114:239--251.

\bibitem{fameau2017non}
Fameau AL, Saint-Jalmes A.
\newblock Non-aqueous foams: Current understanding on the formation and
  stability mechanisms.
\newblock Advances in colloid and interface science. 2017;247:454--464.

\bibitem{exerowa1997foam}
Exerowa D, Kruglyakov PM.
\newblock Foam and foam films: theory, experiment, application.
\newblock Elsevier; 1997.

\bibitem{mcclements2007critical}
Mcclements DJ.
\newblock Critical review of techniques and methodologies for characterization
  of emulsion stability.
\newblock Critical reviews in food science and nutrition. 2007;47(7):611--649.

\bibitem{miralles2014foam}
Miralles V, Selva B, Cantat I, Jullien MC.
\newblock Foam drainage control using thermocapillary stress in a
  two-dimensional microchamber.
\newblock Physical Review Letters. 2014;112(23):238302.

\bibitem{exerowa2018foam}
Exerowa D, Gochev G, Platikanov D, Liggieri L, Miller R.
\newblock Foam Films and Foams: Fundamentals and Applications.
\newblock CRC Press; 2018.

\bibitem{sheludko1967thin}
Sheludko A.
\newblock Thin liquid films.
\newblock Advances in Colloid and Interface Science. 1967;1(4):391--464.

\bibitem{exerowa2008emulsion}
Exerowa D, Gotchev G, Levecke B, Tadros T.
\newblock Emulsion Newton black films stabilized by polymeric surfactants.
\newblock CR Acad Bulgare Sci. 2008;61:455--468.

\bibitem{dippenaar1982destabilization}
Dippenaar A.
\newblock The destabilization of froth by solids. I. The mechanism of film
  rupture.
\newblock International Journal of Mineral Processing. 1982;9(1):1--14.

\bibitem{stubenrauch2003disjoining}
Stubenrauch C, Von~Klitzing R.
\newblock Disjoining pressure in thin liquid foam and emulsion films—new
  concepts and perspectives.
\newblock Journal of Physics: condensed matter. 2003;15(27):R1197.

\bibitem{joye1994asymmetric}
Joye JL, Hirasaki GJ, Miller CA.
\newblock Asymmetric drainage in foam films.
\newblock Langmuir. 1994;10(9):3174--3179.

\bibitem{langevin2015bubble}
Langevin D.
\newblock Bubble coalescence in pure liquids and in surfactant solutions.
\newblock Current Opinion in Colloid \& Interface Science. 2015;20(2):92--97.

\bibitem{chan2011film}
Chan DY, Klaseboer E, Manica R.
\newblock Film drainage and coalescence between deformable drops and bubbles.
\newblock Soft Matter. 2011;7(6):2235--2264.

\bibitem{kannan2018monoclonal}
Kannan A, Shieh IC, Leiske DL, Fuller GG.
\newblock Monoclonal antibody interfaces: dilatation mechanics and bubble
  coalescence.
\newblock Langmuir. 2018;34(2):630--638.

\bibitem{kannan2019linking}
Kannan A, Shieh IC, Fuller GG.
\newblock Linking aggregation and interfacial properties in monoclonal
  antibody-surfactant formulations.
\newblock Journal of colloid and interface science. 2019;550:128--138.

\bibitem{gochev2016chronicles}
Gochev G, Platikanov D, Miller R.
\newblock Chronicles of foam films.
\newblock Advances in Colloid and Interface Science. 2016;233:115--125.

\bibitem{plateau1873statique}
Plateau J.
\newblock Statique exp{\'e}rimentale et th{\'e}orique des liquides soumis aux
  seules forces mol{\'e}culaires. vol.~2.
\newblock Gauthier-Villars; 1873.

\bibitem{isenberg1981soap}
Isenberg C.
\newblock Soap films and bubbles.
\newblock Physics Education. 1981;16(4):218.

\bibitem{boys1890soap}
Boys CV.
\newblock Soap-Bubbles and the Forces which Mould Them.
\newblock Society for Promoting Christian Knowledge; 1890.

\bibitem{worthington1881ii}
Worthington AM.
\newblock II. On pendent drops.
\newblock Proceedings of the Royal Society of London.
  1881;32(212-215):362--377.

\bibitem{edgerton1937studies}
Edgerton HE, Hauser EA, Tucker W.
\newblock Studies in drop formation as revealed by the high-speed motion
  camera.
\newblock Journal of Physical Chemistry. 1937;41(7):1017--1028.

\bibitem{berry2015measurement}
Berry JD, Neeson MJ, Dagastine RR, Chan DY, Tabor RF.
\newblock Measurement of surface and interfacial tension using pendant drop
  tensiometry.
\newblock Journal of colloid and interface science. 2015;454:226--237.

\bibitem{simon1851recherches}
Simon M.
\newblock Recherches sur la capillarit{\'e}.
\newblock In: Annales de Chimie et de Physique. vol.~32; 1851. p.~5.

\bibitem{mysels1990maximum}
Mysels KJ.
\newblock The maximum bubble pressure method of measuring surface tension,
  revisited.
\newblock Colloids and surfaces. 1990;43(2):241--262.

\bibitem{schrodinger1915notiz}
Schr{\"o}dinger E.
\newblock Notiz {\"u}ber den Kapillardruck in Gasblasen.
\newblock Annalen der Physik. 1915;351(3):413--418.

\bibitem{fuller2012complex}
Fuller GG, Vermant J.
\newblock Complex Fluid-Fluid Interfaces: Rheology and Structure.
\newblock Annual Review of Chemical and Biomolecular Engineering.
  2012;3(1):519--543.

\bibitem{marangoni1972principle}
Marangoni C.
\newblock The principle of the surface viscosity of liquids established by mr.
  j. plateau.
\newblock NCim. 1972;5:239--273.

\bibitem{boussinesq1913existence}
Boussinesq J.
\newblock Existence of a superficial viscosity in the thin transition layer
  separating one liquid from another contiguous fluid.
\newblock CR Hehbd Seanc Acad Sci. 1913;156:983--989.

\bibitem{scriven1960dynamics}
Scriven L.
\newblock Dynamics of a fluid interface equation of motion for Newtonian
  surface fluids.
\newblock Chemical Engineering Science. 1960;12(2):98--108.

\bibitem{lunkenheimer1984investigation}
Lunkenheimer K, Hartenstein C, Miller R, Wantke KD.
\newblock Investigation on the method of the radially oscillating bubble.
\newblock Colloids and surfaces. 1984;8(3):271--288.

\bibitem{nagarajan1993measurement}
Nagarajan R, Wasan D.
\newblock Measurement of dynamic interfacial tension by an expanding drop
  tensiometer.
\newblock Journal of colloid and interface science. 1993;159(1):164--173.

\bibitem{rayleigh1882xx}
Rayleigh L.
\newblock XX. On the equilibrium of liquid conducting masses charged with
  electricity.
\newblock The London, Edinburgh, and Dublin Philosophical Magazine and Journal
  of Science. 1882;14(87):184--186.

\bibitem{clayton1923theory}
Clayton W.
\newblock The theory of emulsions and emulsification.
\newblock J. \& A. Churchill; 1923.

\bibitem{wilson1925bursting}
Wilson C, Taylor G; Cambridge~University Press.
\newblock The bursting of soap-bubbles in a uniform electric field.
\newblock Mathematical proceedings of the Cambridge philosophical society.
  1925;22(5):728--730.

\bibitem{dewar1919soap}
Dewar J.
\newblock Soap bubbles of long duration.
\newblock Journal of the Franklin Institute. 1919;188(6):713--749.

\bibitem{derjaguin1939anomalous}
Derjaguin B, Kussakov M.
\newblock Anomalous properties of thin polymolecular films.
\newblock Acta Physicochim URSS. 1939;10(1):25--44.

\bibitem{rehbinder1930stabilisierende}
Rehbinder P, Wenstr{\"o}m E.
\newblock Stabilisierende Wirkung von Adsorptionsschichten
  grenzfl{\"a}chenaktiver Stoffe auf disperse Systeme. II. Stabilit{\"a}t von
  Blasen und Tropfen an Trennungsfl{\"a}chen.
\newblock Kolloid-Zeitschrift. 1930;53(2):145--158.

\bibitem{charles1960coalescence}
Charles GE, Mason SG.
\newblock The coalescence of liquid drops with flat liquid/liquid interfaces.
\newblock Journal of Colloid Science. 1960;15(3):236--267.

\bibitem{frostad2016dynamic}
Frostad JM, Tammaro D, Santollani L, de~Araujo SB, Fuller GG.
\newblock Dynamic fluid-film interferometry as a predictor of bulk foam
  properties.
\newblock Soft matter. 2016;12(46):9266--9279.

\bibitem{chandran2016impact}
Chandran~Suja V, Frostad J, Fuller G.
\newblock Impact of compressibility on the control of bubble-pressure
  tensiometers.
\newblock Langmuir. 2016;32(46):12031--12038.

\bibitem{suja2020symmetry}
Chandran~Suja V, Hadidi A, Kannan A, Fuller GG.
\newblock Symmetry breaking and chaos in evaporation driven Marangoni flows
  over bubbles.
\newblock arXiv preprint arXiv:200409752. 2020;.

\bibitem{allan1961approach}
Allan RS, Charles G, Mason S.
\newblock The approach of gas bubbles to a gas/liquid interface.
\newblock Journal of Colloid Science. 1961;16(2):150--165.

\bibitem{zhang2015partial}
Zhang F, Thoraval MJ, Thoroddsen ST, Taborek P.
\newblock Partial coalescence from bubbles to drops.
\newblock Journal of Fluid Mechanics. 2015;782:209--239.

\bibitem{feng2016dynamics}
Feng J, Muradoglu M, Kim H, Ault JT, Stone HA.
\newblock Dynamics of a bubble bouncing at a liquid/liquid/gas interface.
\newblock Journal of Fluid Mechanics. 2016;807:324--352.

\bibitem{krzan2003pulsation}
Krzan M, Lunkenheimer K, Malysa K.
\newblock Pulsation and bouncing of a bubble prior to rupture and/or foam film
  formation.
\newblock Langmuir. 2003;19(17):6586--6589.

\bibitem{menesses2019surfactant}
Menesses M, Roch{\'e} M, Royon L, Bird JC.
\newblock Surfactant-free persistence of surface bubbles in a volatile liquid.
\newblock Physical Review Fluids. 2019;4(10):100506.

\bibitem{sunol2010rise}
Su{\~n}ol F, Gonz{\'a}lez-Cinca R.
\newblock Rise, bouncing and coalescence of bubbles impacting at a free
  surface.
\newblock Colloids and Surfaces A: Physicochemical and Engineering Aspects.
  2010;365(1-3):36--42.

\bibitem{sett2013gravitational}
Sett S, Sinha-Ray S, Yarin A.
\newblock Gravitational drainage of foam films.
\newblock Langmuir. 2013;29(16):4934--4947.

\bibitem{paulsen2014coalescence}
Paulsen JD, Carmigniani R, Kannan A, Burton JC, Nagel SR.
\newblock Coalescence of bubbles and drops in an outer fluid.
\newblock Nature communications. 2014;5(1):1--7.

\bibitem{bluteau2017water}
Bluteau L, Bourrel M, Passade-Boupat N, Talini L, Verneuil E, Lequeux F.
\newblock Water film squeezed between oil and solid: drainage towards
  stabilization by disjoining pressure.
\newblock Soft matter. 2017;13(7):1384--1395.

\bibitem{blanchette2006partial}
Blanchette F, Bigioni TP.
\newblock Partial coalescence of drops at liquid interfaces.
\newblock Nature Physics. 2006;2(4):254--257.

\bibitem{mackay1963gravity}
MacKay G, Mason S.
\newblock The gravity approach and coalescence of fluid drops at liquid
  interfaces.
\newblock The Canadian Journal of Chemical Engineering. 1963;41(5):203--212.

\bibitem{gillespie1956coalescence}
Gillespie T, Rideal EK.
\newblock The coalescence of drops at an oil-water interface.
\newblock Transactions of the Faraday Society. 1956;52:173--183.

\bibitem{aarts2008droplet}
Aarts DG, Lekkerkerker HN.
\newblock Droplet coalescence: drainage, film rupture and neck growth in
  ultralow interfacial tension systems.
\newblock Journal of fluid mechanics. 2008;606:275--294.

\bibitem{klaseboer2000film}
Klaseboer E, Chevaillier JP, Gourdon C, Masbernat O.
\newblock Film drainage between colliding drops at constant approach velocity:
  experiments and modeling.
\newblock Journal of colloid and interface science. 2000;229(1):274--285.

\bibitem{vakarelski2010dynamic}
Vakarelski IU, Manica R, Tang X, O’Shea SJ, Stevens GW, Grieser F, et~al.
\newblock Dynamic interactions between microbubbles in water.
\newblock Proceedings of the National Academy of Sciences.
  2010;107(25):11177--11182.

\bibitem{zheng1983laboratory}
Zheng Q, Klemas V, Hsu YH.
\newblock Laboratory measurement of water surface bubble life time.
\newblock Journal of Geophysical Research: Oceans. 1983;88(C1):701--706.

\bibitem{vitry2019controlling}
Vitry Y, Dorbolo S, Vermant J, Scheid B.
\newblock Controlling the lifetime of antibubbles.
\newblock Advances in colloid and interface science. 2019;.

\bibitem{poulain2018ageing}
Poulain S, Villermaux E, Bourouiba L.
\newblock Ageing and burst of surface bubbles.
\newblock Journal of Fluid Mechanics. 2018;851:636--671.

\bibitem{Kabalnov1998coalescence}
Kabalnov AS.
\newblock Chapter 7 - Coalescence in Emulsions.
\newblock In: Modern Aspects of Emulsion Science. The Royal Society of
  Chemistry; 1998. p. 205--260.

\bibitem{Deminiere1999cell}
Deminiere B, Colin A, Leal-Calderon F, Muzy JF, Bibette J.
\newblock Cell Growth in a 3D Cellular System Undergoing Coalescence.
\newblock Phys Rev Lett. 1999 Jan;82:229--232.

\bibitem{tobin2011public}
Tobin S, Meagher A, Bulfin B, M{\"o}bius M, Hutzler S.
\newblock A public study of the lifetime distribution of soap films.
\newblock American Journal of Physics. 2011;79(8):819--824.

\bibitem{ghosh2004coalescence}
Ghosh P.
\newblock Coalescence of air bubbles at air--water interface.
\newblock Chemical Engineering Research and Design. 2004;82(7):849--854.

\bibitem{lhuissier2012bursting}
Lhuissier H, Villermaux E.
\newblock Bursting bubble aerosols.
\newblock Journal of Fluid Mechanics. 2012;696:5--44.

\bibitem{ghosh2002analysis}
Ghosh P, Juvekar V.
\newblock Analysis of the drop rest phenomenon.
\newblock Chemical Engineering Research and Design. 2002;80(7):715--728.

\bibitem{sornette2006critical}
Sornette D.
\newblock Critical phenomena in natural sciences: chaos, fractals,
  selforganization and disorder: concepts and tools.
\newblock Springer Science \& Business Media; 2006.

\bibitem{moon1996expectation}
Moon TK.
\newblock The expectation-maximization algorithm.
\newblock IEEE Signal processing magazine. 1996;13(6):47--60.

\bibitem{bhamla2016instability}
Bhamla MS, Chai C, Rabiah NI, Frostad JM, Fuller GG.
\newblock Instability and breakup of model tear films.
\newblock Investigative ophthalmology \& visual science. 2016;57(3):949--958.

\bibitem{hermans2015lung}
Hermans E, Bhamla MS, Kao P, Fuller GG, Vermant J.
\newblock Lung surfactants and different contributions to thin film stability.
\newblock Soft matter. 2015;11(41):8048--8057.

\bibitem{shukla2006non}
Shukla R, Udupa D, Das N, Mantravadi MV.
\newblock Non-destructive thickness measurement of dichromated gelatin films
  deposited on glass plates.
\newblock Optics \& Laser Technology. 2006;38(7):552--557.

\bibitem{greco1994measuring}
Greco V, Lemmi C, Ledesma S, Molesini G, Puccioni G, Quercioli F.
\newblock Measuring soap black films by phase shifting interferometry.
\newblock Measurement Science and Technology. 1994;5(8):900.

\bibitem{habibi2007langmuir}
Habibi Y, Foulon L, Agui{\'e}-B{\'e}ghin V, Molinari M, Douillard R.
\newblock Langmuir--Blodgett films of cellulose nanocrystals: Preparation and
  characterization.
\newblock Journal of Colloid and Interface Science. 2007;316(2):388--397.

\bibitem{zhang2015domain}
Zhang Y, Sharma V.
\newblock Domain expansion dynamics in stratifying foam films: experiments.
\newblock Soft matter. 2015;11(22):4408--4417.

\bibitem{zhang2016nanoscopic}
Zhang Y, Yilixiati S, Pearsall C, Sharma V.
\newblock Nanoscopic terraces, mesas, and ridges in freely standing thin films
  sculpted by supramolecular oscillatory surface forces.
\newblock ACS nano. 2016;10(4):4678--4683.

\bibitem{lv2012spatial}
Lv W, Zhou H, Lou C, Zhu J.
\newblock Spatial and temporal film thickness measurement of a soap bubble
  based on large lateral shearing displacement interferometry.
\newblock Applied optics. 2012;51(36):8863--8872.

\bibitem{vannoni2013measuring}
Vannoni M, Sordini A, Gabrieli R, Melozzi M, Molesini G.
\newblock Measuring the thickness of soap bubbles with phase-shift
  interferometry.
\newblock Optics express. 2013;21(17):19657--19667.

\bibitem{beltramo2016simple}
Beltramo PJ, Vermant J.
\newblock Simple Optical Imaging of Nanoscale Features in Free-Standing Films.
\newblock ACS omega. 2016;1(3):363--370.

\bibitem{beltramo2016millimeter}
Beltramo PJ, Van~Hooghten R, Vermant J.
\newblock Millimeter-area, free standing, phospholipid bilayers.
\newblock Soft Matter. 2016;12(19):4324--4331.

\bibitem{kralchevsky1990interaction}
Kralchevsky P, Nikolov A, Wasan D, Ivanov I.
\newblock Interaction of colloid particles in thinning foam films. Formation
  and expansion of dark spots.
\newblock Langmuir. 1990;6:1180.

\bibitem{nikolov1989ordered}
Nikolov A, Wasan D.
\newblock Ordered micelle structuring in thin films formed from anionic
  surfactant solutions. I. Experimental.
\newblock J Colloid Interface Sci. 1989;133(1):1--12.

\bibitem{hendrix2012spatiotemporal}
Hendrix MH, Manica R, Klaseboer E, Chan DY, Ohl CD.
\newblock Spatiotemporal evolution of thin liquid films during impact of water
  bubbles on glass on a micrometer to nanometer scale.
\newblock Physical review letters. 2012;108(24):247803.

\bibitem{czakaj2020viscoelastic}
Czakaj A, Kannan A, Wi{\'s}niewska A, Grze{\'s} G, Krzan M, Warszy{\'n}ski P,
  et~al.
\newblock Viscoelastic interfaces comprising of cellulose nanocrystals and
  lauroyl ethyl arginate for enhanced foam stability.
\newblock Soft Matter. 2020;.

\bibitem{rodriguez2019evaporation}
Rodr{\'\i}guez-Hakim M, Barakat JM, Shi X, Shaqfeh ES, Fuller GG.
\newblock Evaporation-driven solutocapillary flow of thin liquid films over
  curved substrates.
\newblock Physical Review Fluids. 2019;4(3):034002.

\bibitem{rabiah2019influence}
Rabiah NI, Scales CW, Fuller GG.
\newblock The influence of protein deposition on contact lens tear film
  stability.
\newblock Colloids and Surfaces B: Biointerfaces. 2019;180:229--236.

\bibitem{nierstrasz1999marginal}
Nierstrasz VA, Frens G.
\newblock Marginal regeneration and the Marangoni effect.
\newblock Journal of colloid and interface science. 1999;215(1):28--35.

\bibitem{lee2014surfactant}
Lee J, Nikolov A, Wasan D.
\newblock Surfactant micelles containing solubilized oil decrease foam film
  thickness stability.
\newblock Journal of colloid and interface science. 2014;415:18--25.

\bibitem{rabiah2020understanding}
Rabiah NI, Sato Y, Kannan A, Kress W, Straube F, Fuller GG.
\newblock Understanding the adsorption and potential tear film stability
  properties of recombinant human lubricin and bovine submaxillary mucins in an
  in vitro tear film model.
\newblock Colloids and Surfaces B: Biointerfaces. 2020;p. 111257.

\bibitem{kitagawa2013thin}
Kitagawa K.
\newblock Thin-film thickness profile measurement by three-wavelength
  interference color analysis.
\newblock Applied optics. 2013;52(10):1998--2007.

\bibitem{doane1989instrument}
Doane MG.
\newblock An instrument for in vivo tear film interferometry.
\newblock Optometry and vision science: official publication of the American
  Academy of Optometry. 1989;66(6):383--388.

\bibitem{blodgett1952step}
Blodgett KB. Step gauge for measuring thickness of thin films; 1952.
\newblock US Patent 2,587,282.

\bibitem{hecht2017optics}
Hecht E.
\newblock Optics.
\newblock Pearson Education, Incorporated; 2017.

\bibitem{karakashev2008effect}
Karakashev SI, Manev ED, Tsekov R, Nguyen AV.
\newblock Effect of ionic surfactants on drainage and equilibrium thickness of
  emulsion films.
\newblock Journal of colloid and interface science. 2008;318(2):358--364.

\bibitem{dong2019laser}
Dong T, Weheliye WH, Angeli P.
\newblock Laser induced fluorescence studies on the distribution of surfactants
  during drop/interface coalescence.
\newblock Physics of Fluids. 2019;31(1):012106.

\bibitem{bordoloi2012effect}
Bordoloi AD, Longmire EK.
\newblock Effect of neighboring perturbations on drop coalescence at an
  interface.
\newblock Physics of Fluids. 2012;24(6):062106.

\bibitem{suja2020hyperspectral}
Suja VC, Sentmanat J, Hofmann G, Scales C, Fuller GG.
\newblock Hyperspectral imaging for dynamic thin film interferometry.
\newblock Scientific reports. 2020;10(1):11378.

\bibitem{mandracchia2019quantitative}
Mandracchia B, Wang Z, Ferraro V, Villone MM, Di~Maio E, Maffettone PL, et~al.
\newblock Quantitative imaging of the complexity in liquid bubbles’ evolution
  reveals the dynamics of film retraction.
\newblock Light: Science \& Applications. 2019;8(1):1--12.

\bibitem{lin2016interfacial}
Lin GL, Pathak JA, Kim DH, Carlson M, Riguero V, Kim YJ, et~al.
\newblock Interfacial dilatational deformation accelerates particle formation
  in monoclonal antibody solutions.
\newblock Soft Matter. 2016;12(14):3293--3302.

\bibitem{lin2018influence}
Lin G, Frostad JM, Fuller GG.
\newblock Influence of interfacial elasticity on liquid entrainment in thin
  foam films.
\newblock Physical Review Fluids. 2018;3(11):114001.

\bibitem{kannan2020surfactant}
Kannan A, Hristov P, Li J, Zawala J, Gao P, Fuller GG.
\newblock Surfactant-laden bubble dynamics under porous polymer films.
\newblock Journal of Colloid and Interface Science. 2020;.

\bibitem{rodriguez2020asphaltene}
Rodr{\'\i}guez-Hakim M, Anand S, Tajuelo J, Yao Z, Kannan A, Fuller GG.
\newblock Asphaltene-induced spontaneous emulsification: Effects of interfacial
  co-adsorption and viscoelasticity.
\newblock Journal of Rheology. 2020;64(4):799--816.

\bibitem{nagel2017drop}
Nagel M, Tervoort TA, Vermant J.
\newblock From drop-shape analysis to stress-fitting elastometry.
\newblock Advances in colloid and interface science. 2017;247:33--51.

\bibitem{jaensson2018tensiometry}
Jaensson N, Vermant J.
\newblock Tensiometry and rheology of complex interfaces.
\newblock Current Opinion in Colloid \& Interface Science. 2018;.

\bibitem{kotula2015regular}
Kotula AP, Anna SL.
\newblock Regular perturbation analysis of small amplitude oscillatory
  dilatation of an interface in a capillary pressure tensiometer.
\newblock Journal of Rheology. 2015;59(1):85--117.

\bibitem{wilde2004proteins}
Wilde P, Mackie A, Husband F, Gunning P, Morris V.
\newblock Proteins and emulsifiers at liquid interfaces.
\newblock Advances in Colloid and Interface Science. 2004;108-109:63 -- 71.
\newblock Emulsions, From Fundamentals to Practical Applications.

\bibitem{sanchez2018dynamic}
Sánchez-Puga P, Tajuelo J, Pastor J, Rubio M.
\newblock Dynamic Measurements with the Bicone Interfacial Shear Rheometer:
  Numerical Bench-Marking of Flow Field-Based Data Processing.
\newblock Colloids and Interfaces. 2018 Dec;2(4):69.

\bibitem{carvajal2011mechanics}
Carvajal D, Laprade EJ, Henderson KJ, Shull KR.
\newblock Mechanics of pendant drops and axisymmetric membranes.
\newblock Soft Matter. 2011;7:10508--10519.

\bibitem{danov2015capillary}
Danov KD, Stanimirova RD, Kralchevsky PA, Marinova KG, Alexandrov NA, Stoyanov
  SD, et~al.
\newblock Capillary meniscus dynamometry - A method for determining the surface
  tension of drops and bubbles with isotropic and anisotropic surface stress
  distributions.
\newblock Journal of Colloid and Interface Science. 2015;440:168 -- 178.

\bibitem{leal2007advanced}
Leal LG.
\newblock {Advanced transport phenomena: fluid mechanics and convective
  transport processes}.
\newblock Cambridge University Press; 2007.

\bibitem{berg2012introduction}
Berg JC.
\newblock {An Introduction to Interfaces and Colloids: The Bridge to
  Nanoscience}.
\newblock World Scientific Publishing Co.; 2012.

\bibitem{murphy1966thermodynamics}
Murphy CL.
\newblock Thermodynamics of low tension and highly curved interfaces [Ph.D.
  thesis].
\newblock University of Minnesota; 1966.

\bibitem{leermakers2006symmetric}
Leermakers FAM, Barneveld PA, Sprakel J, Besseling NAM.
\newblock {Symmetric Liquid-Liquid Interface with a Nonzero Spontaneous
  Curvature}.
\newblock Phys Rev Lett. 2006;97(066103):1--4.

\bibitem{evans1994hidden}
Evans E, Yeung A.
\newblock {Hidden dynamics in rapid changes of bilayer shape}.
\newblock Chem Phys Lipids. 1994;73:39--56.

\bibitem{helfrich1973elastic}
Helfrich W.
\newblock {Elastic Properties of Lipid Bilayers: Theory and Possible
  Experiments}.
\newblock Z Naturforsch. 1973;28c:693--703.

\bibitem{helfrich1989bending}
Zhong-can OY, Helfrich W.
\newblock {Bending energy of vesicle membranes: General expressions for the
  first, second, and third variation of the shape energy and applications to
  spheres and cylinders}.
\newblock Phys Rev A. 1989;39(10):5280--5288.

\bibitem{gekle2017theory}
Guckenberger A, Gekle s.
\newblock {Theory and algorithms to compute Helfrich bending forces: a review}.
\newblock J Phys: Condens Matter. 2017;29:203001.

\bibitem{lopez1998direct}
Lopez J, Hirsa A.
\newblock Direct determination of the dependence of the surface shear and
  dilatational viscosities on the thermodynamic state of the interface:
  Theoretical foundations.
\newblock Journal of colloid and interface science. 1998;206(1):231--239.

\bibitem{pepicelli2017characterization}
Pepicelli M, Verwijlen T, Tervoort TA, Vermant J.
\newblock Characterization and modelling of Langmuir interfaces with finite
  elasticity.
\newblock Soft Matter. 2017;13:5977--5990.

\bibitem{rotenberg1983determination}
Rotenberg Y, Boruvka L, Neumann A.
\newblock Determination of surface tension and contact angle from the shapes of
  axisymmetric fluid interfaces.
\newblock Journal of colloid and interface science. 1983;93(1):169--183.

\bibitem{deGennes2004capillarity}
de~Gennes PG, Brochard-Wyart F, Quere D.
\newblock Capillarity and Wetting Phenomena.
\newblock Springer; 2004.

\bibitem{alvarez2009non}
Alvarez NJ, Walker LM, Anna SL.
\newblock A non-gradient based algorithm for the determination of surface
  tension from a pendant drop: Application to low Bond number drop shapes.
\newblock Journal of colloid and interface science. 2009;333(2):557--562.

\bibitem{andreas2002boundary}
Andreas J, Hauser E, Tucker W.
\newblock Boundary tension by pendant drops.
\newblock The Journal of Physical Chemistry. 1938;42(8):1001--1019.

\bibitem{juuza1997pendant}
Juuza J.
\newblock The pendant drop method of surface tension measurement: equation
  interpolating the shape factor tables for several selected planes.
\newblock Czechoslovak Journal of Physics. 1997;47(3):351--357.

\bibitem{stauffer1965measurement}
Stauffer CE.
\newblock The measurement of surface tension by the pendant drop technique.
\newblock The journal of physical chemistry. 1965;69(6):1933--1938.

\bibitem{neeson2014compound}
Neeson MJ, Chan DY, Tabor RF.
\newblock Compound pendant drop tensiometry for interfacial tension measurement
  at zero Bond number.
\newblock Langmuir. 2014;30(51):15388--15391.

\bibitem{buzzacchi2006dynamic}
Buzzacchi M, Schmiedel P, von Rybinski W.
\newblock Dynamic surface tension of surfactant systems and its relation to
  foam formation and liquid film drainage on solid surfaces.
\newblock Colloids and Surfaces A: Physicochemical and Engineering Aspects.
  2006;273(1-3):47--54.

\bibitem{fainerman1998maximum}
Fainerman V, Miller R.
\newblock The maximum bubble pressure tensiometry.
\newblock In: Studies in Interface Science. vol.~6. Elsevier; 1998. p.
  279--326.

\bibitem{bendure1971dynamic}
Bendure RL.
\newblock Dynamic surface tension determination with the maximum bubble
  pressure method.
\newblock Journal of Colloid and Interface Science. 1971;35(2):238--248.

\bibitem{javadi2012fast}
Javadi A, Kragel J, Makievski AV, Kovalchuk VI, Kovalchuk NM, Mucic N, et~al.
\newblock Fast dynamic interfacial tension measurements and dilational rheology
  of interfacial layers by using the capillary pressure technique.
\newblock Colloids and Surfaces A: Physicochemical and Engineering Aspects.
  2012;407:159 -- 168.

\bibitem{alvarez2010microtensiometer}
Alvarez N, Walker L, Anna S.
\newblock A Microtensiometer To Probe the Effect of Radius of Curvature on
  Surfactant Transport to a Spherical Interface.
\newblock Langmuir. 2010 08;26:13310--13319.

\bibitem{kazakov2008dilatational}
Kazakov V, Fainerman VB, Kondratenko PG, Elin AF, Sinyachenko OV, Miller R.
\newblock Dilational rheology of serum albumin and blood serum solutions as
  studied by oscillating drop tensiometry.
\newblock Colloids and surfaces B, Biointerfaces. 2008 04;62:77--82.

\bibitem{manga2016measurements}
Manga M, Hunter T, Cayre O, York D, Reichert M, Anna S, et~al.
\newblock Measurements of Submicron Particle Adsorption and Particle Film
  Elasticity at Oil-Water Interfaces.
\newblock Langmuir : the ACS journal of surfaces and colloids. 2016 04;32.

\bibitem{alvarez2012interfacial}
Alvarez NJ, Anna SL, Saigal T, Tilton RD, Walker LM.
\newblock Interfacial dynamics and rheology of polymer-grafted nanoparticles at
  air--water and xylene--water interfaces.
\newblock Langmuir. 2012;28(21):8052--8063.

\bibitem{russev2008instrument}
Russev S, Alexandrov N, Marinova K, Danov K, Denkov N, Lyutov L, et~al.
\newblock Instrument and methods for surface dilatational rheology
  measurements.
\newblock The Review of scientific instruments. 2008 11;79:104102.

\bibitem{liggieri2002measurement}
Liggieri L, Attolini V, Ferrari M, Ravera F.
\newblock Measurement of the surface dilational viscoelasticity of adsorbed
  layers with a capillary pressure tensiometer.
\newblock Journal of colloid and interface science. 2002;255(2):225--235.

\bibitem{ravera2010interfacial}
Ravera F, Loglio G, Kovalchuk VI.
\newblock Interfacial dilational rheology by oscillating bubble/drop methods.
\newblock Current Opinion in Colloid \& Interface Science. 2010;15(4):217 --
  228.

\bibitem{narayan2018removing}
Narayan S, Moravec DB, Hauser BG, Dallas AJ, Dutcher CS.
\newblock Removing Water from Diesel Fuel: Understanding the Impact of Droplet
  Size on Dynamic Interfacial Tension of Water-in-Fuel Emulsions.
\newblock Energy \& Fuels. 2018;32(7):7326--7337.

\bibitem{chen2020size}
Chen Y, Dutcher CS.
\newblock Size dependent droplet interfacial tension and surfactant transport
  in liquid-liquid systems, with applications in shipboard oily bilgewater
  emulsions.
\newblock Soft Matter. 2020;16:2994--3004.

\bibitem{pepicelli2019surface}
Pepicelli M, Jaensson N, TregouÃ«t C, Schroyen B, Alicke A, Tervoort T,
  et~al.
\newblock Surface viscoelasticity in model polymer multilayers: From planar
  interfaces to rising bubbles.
\newblock Journal of Rheology. 2019;63(5):815--828.

\bibitem{freer2004relaxation}
Freer EM, Radke CJ.
\newblock {Relaxation of asphaltenes at the toluene/water interface: diffusion
  exchange and surface rearrangement}.
\newblock The Journal of Adhesion. 2004;80:481--496.

\bibitem{loglio2005perturbation}
Loglio G, Pandolfini P, Miller R, Makievski AV, KrÃ€gel J, Ravera F, et~al.
\newblock Perturbation-response relationship in liquid interfacial systems:
  non-linearity assessment by frequency-domain analysis.
\newblock Colloids and Surfaces A: Physicochemical and Engineering Aspects.
  2005;261(1):57 -- 63.

\bibitem{grassia2002foam}
Grassia P, Neethling S, Cilliers J.
\newblock Foam drainage on a sloping weir.
\newblock The European Physical Journal E. 2002;8(1):517--529.

\bibitem{stevenson2006wetness}
Stevenson P.
\newblock The wetness of a rising foam.
\newblock Industrial \& engineering chemistry research. 2006;45(2):803--807.

\bibitem{de1990rheology}
de~Vries AS, Wit K, et~al.
\newblock Rheology of gas/water foam in the quality range relevant to steam
  foam.
\newblock SPE Reservoir Engineering. 1990;5(02):185--192.

\bibitem{simjoo2013foam}
Simjoo M, Rezaei T, Andrianov A, Zitha P.
\newblock Foam stability in the presence of oil: effect of surfactant
  concentration and oil type.
\newblock Colloids and Surfaces A: Physicochemical and Engineering Aspects.
  2013;438:148--158.

\bibitem{binks2010non}
Binks B, Davies C, Fletcher P, Sharp E.
\newblock Non-aqueous foams in lubricating oil systems.
\newblock Colloids and Surfaces A: Physicochemical and Engineering Aspects.
  2010;360(1-3):198--204.

\bibitem{bhamla2014influence}
Bhamla MS, Giacomin CE, Balemans C, Fuller GG.
\newblock Influence of interfacial rheology on drainage from curved surfaces.
\newblock Soft Matter. 2014;10(36):6917--6925.

\bibitem{garrett1979effect}
Garrett PR.
\newblock The effect of polytetrafluoroethylene particles on the foamability of
  aqueous surfactant solutions.
\newblock Journal of colloid and interface science. 1979;69(1):107--121.

\bibitem{langevin2000influence}
Langevin D.
\newblock Influence of interfacial rheology on foam and emulsion properties.
\newblock Advances in Colloid and Interface Science. 2000;88(1-2):209--222.

\bibitem{tuinier1996transient}
Tuinier R, Bisperink CG, van~den Berg C, Prins A.
\newblock Transient foaming behavior of aqueous alcohol solutions as related to
  their dilational surface properties.
\newblock Journal of colloid and interface science. 1996;179(2):327--334.

\bibitem{lorenceau2020lifetime}
Lorenceau E, Rouyer F.
\newblock Lifetime of a single bubble on the surface of a water and ethanol
  bath.
\newblock Physical Review Fluids. 2020;5(6):063603.

\bibitem{shi2020oscillatory}
Shi X, Fuller GG, Shaqfeh ES.
\newblock Oscillatory spontaneous dimpling in evaporating curved thin films.
\newblock Journal of Fluid Mechanics. 2020;889.

\bibitem{velev1993spontaneous}
Velev OD, Gurkov TD, Borwankar RP.
\newblock Spontaneous cyclic dimpling in emulsion films due to surfactant mass
  transfer between the phases.
\newblock Journal of colloid and interface science. 1993;159:497--497.

\bibitem{muller2007experimental}
M{\"u}ller F, Kornek U, Stannarius R.
\newblock Experimental study of the bursting of inviscid bubbles.
\newblock Physical Review E. 2007;75(6):065302.

\bibitem{pandit1990hydrodynamics}
Pandit A, Davidson J.
\newblock Hydrodynamics of the rupture of thin liquid films.
\newblock Journal of Fluid Mechanics. 1990;212:11--24.

\bibitem{muller2009comparison}
M{\"u}ller F, Stannarius R.
\newblock Comparison of the rupture dynamics of smectic bubbles and soap
  bubbles.
\newblock Liquid Crystals. 2009;36(2):133--145.

\bibitem{robinson2001observations}
Robinson ND, Steen PH.
\newblock Observations of singularity formation during the capillary collapse
  and bubble pinch-off of a soap film bridge.
\newblock Journal of colloid and interface science. 2001;241(2):448--458.

\bibitem{tammaro2018elasticity}
Tammaro D, Pasquino R, Villone MM, D’Avino G, Ferraro V, Di~Maio E, et~al.
\newblock Elasticity in bubble rupture.
\newblock Langmuir. 2018;34(19):5646--5654.

\bibitem{sabadini2014elasticity}
Sabadini E, Ungarato RF, Miranda PB.
\newblock The elasticity of soap bubbles containing wormlike micelles.
\newblock Langmuir. 2014;30(3):727--732.

\bibitem{debregeas1998life}
Debr{\'e}geas Gd, De~Gennes PG, Brochard-Wyart F.
\newblock The life and death of" bare" viscous bubbles.
\newblock Science. 1998;279(5357):1704--1707.

\bibitem{poulain2018biosurfactants}
Poulain S, Bourouiba L.
\newblock Biosurfactants change the thinning of contaminated bubbles at
  bacteria-laden water interfaces.
\newblock Physical Review Letters. 2018;121(20):204502.

\bibitem{bico2015cracks}
Bico J.
\newblock Cracks in bursting soap films.
\newblock Journal of Fluid Mechanics. 2015;778:1--4.

\bibitem{deVries1958foam}
De~Vries A.
\newblock Foam stability: Part IV. Kinetics and activation energy of film
  rupture.
\newblock Recueil des Travaux Chimiques des Pays-Bas. 1958;77(4):383--399.

\bibitem{Langevin2019coalescence}
Langevin D.
\newblock Coalescence in foams and emulsions: Similarities and differences.
\newblock Current Opinion in Colloid \& Interface Science. 2019;44:23 -- 31.
\newblock Memorial Volume.

\bibitem{Hildebrand1941emulsion}
Hildebrand JH.
\newblock Emulsion Type.
\newblock The Journal of Physical Chemistry. 1941;45(8):1303--1305.

\bibitem{Traykov1977hydrodynamics}
Traykov TT, Manev ED, Ivanov IB.
\newblock Hydrodynamics of thin liquid films. Experimental investigation of the
  effect of surfactants on the drainage of emulsion films.
\newblock International Journal of Multiphase Flow. 1977;3(5):485 -- 494.

\bibitem{Ivanov1997stability}
Ivanov IB, Kralchevsky PA.
\newblock Stability of emulsions under equilibrium and dynamic conditions.
\newblock Colloids and surfaces A, Physicochemical and engineering aspects.
  1997;128(1-3):155--175.

\bibitem{pickering1907emulsions}
Pickering SU.
\newblock Cxcvi.—Emulsions.
\newblock Journal of the Chemical Society, Transactions. 1907;91:2001--2021.

\bibitem{monegier2015influence}
Du~Sorbier QM, Aimable A, Pagnoux C.
\newblock Influence of the electrostatic interactions in a Pickering emulsion
  polymerization for the synthesis of silica--polystyrene hybrid nanoparticles.
\newblock Journal of colloid and interface science. 2015;448:306--314.

\bibitem{menon1988characterization}
Menon V, Wasan D.
\newblock Characterization of oil—water interfaces containing finely divided
  solids with applications to the coalescence of water-in-oil Emulsions: A
  review.
\newblock Colloids and surfaces. 1988;29(1):7--27.

\bibitem{binks2002solid}
Binks BP, Clint JH.
\newblock Solid wettability from surface energy components: relevance to
  Pickering emulsions.
\newblock Langmuir. 2002;18(4):1270--1273.

\bibitem{aveyard2003emulsions}
Aveyard R, Binks BP, Clint JH.
\newblock Emulsions stabilised solely by colloidal particles.
\newblock Advances in Colloid and Interface Science. 2003;100:503--546.

\bibitem{yang2017overview}
Yang Y, Fang Z, Chen X, Zhang W, Xie Y, Chen Y, et~al.
\newblock An overview of Pickering emulsions: solid-particle materials,
  classification, morphology, and applications.
\newblock Frontiers in pharmacology. 2017;8:287.

\bibitem{chen2010novel}
Chen W, Liu X, Liu Y, Kim HI.
\newblock Novel synthesis of self-assembled CNT microcapsules by O/W Pickering
  emulsions.
\newblock Materials Letters. 2010;64(23):2589--2592.

\bibitem{williams2015inorganic}
Williams M, Olland B, Armes SP, Verstraete P, Smets J.
\newblock Inorganic/organic hybrid microcapsules: Melamine formaldehyde-coated
  Laponite-based Pickering emulsions.
\newblock Journal of colloid and interface science. 2015;460:71--80.

\bibitem{frelichowska2009pickering}
Frelichowska J, Bolzinger MA, Valour JP, Mouaziz H, Pelletier J, Chevalier Y.
\newblock Pickering w/o emulsions: drug release and topical delivery.
\newblock International journal of pharmaceutics. 2009;368(1-2):7--15.

\bibitem{Kashchiev1980nucleation}
Kashchiev D, Exerowa D.
\newblock Nucleation mechanism of rupture of Newtonian black films. I. Theory.
\newblock Journal of Colloid and Interface Science. 1980;77(2):501--511.

\bibitem{Kabalnov1996macroemulsion}
Kabalnov A, Wennerstr{\"o}m H.
\newblock Macroemulsion stability: the oriented wedge theory revisited.
\newblock Langmuir. 1996;12(2):276--292.

\bibitem{deGennes2001some}
de~Gennes PG.
\newblock Some remarks on coalescence in emulsions or foams.
\newblock Chemical engineering science. 2001;56(19):5449--5450.

\bibitem{Frostad2013scaling}
Frostad J, Walter J, Leal L.
\newblock A scaling relation for the capillary-pressure driven drainage of thin
  films.
\newblock Physics of Fluids. 2013;25(5):052108.

\bibitem{Kavehpour2015coalescence}
Kavehpour HP.
\newblock Coalescence of drops.
\newblock Annual Review of Fluid Mechanics. 2015;47:245--268.

\bibitem{Fullana1999stability}
Fullana JM, Zaleski S.
\newblock Stability of a growing end rim in a liquid sheet of uniform
  thickness.
\newblock Physics of Fluids. 1999;11(5):952--954.

\bibitem{Menchaca2001coalescence}
Menchaca-Rocha A, Mart{\'\i}nez-D{\'a}valos A, Nunez R, Popinet S, Zaleski S.
\newblock Coalescence of liquid drops by surface tension.
\newblock Physical Review E. 2001;63(4):046309.

\bibitem{Duchemin2002inviscid}
Duchemin L, Eggers J, Josserand C.
\newblock Inviscid coalescence of drops.
\newblock arXiv preprint physics/0212075. 2002;.

\bibitem{Aarts2005hydrodynamics}
Aarts DG, Lekkerkerker HN, Guo H, Wegdam GH, Bonn D.
\newblock Hydrodynamics of droplet coalescence.
\newblock Physical review letters. 2005;95(16):164503.

\bibitem{Thoroddsen2005coalescence}
Thoroddsen S, Takehara K, Etoh T.
\newblock The coalescence speed of a pendent and a sessile drop.
\newblock Journal of Fluid Mechanics. 2005;527:85.

\bibitem{Aryafar2006inertia}
Aryafar H, Lukyanets A, Kavehpour H.
\newblock Inertia-dominated coalescence of drops.
\newblock Applied Mathematics Research eXpress. 2006;2006.

\bibitem{eggers1999coalescence}
Eggers J, Lister JR, Stone HA.
\newblock Coalescence of liquid drops.
\newblock Journal of Fluid Mechanics. 1999;401:293--310.

\bibitem{Zhang2019initial}
Zhang Q, Jiang X, Brunello D, Fu T, Zhu C, Ma Y, et~al.
\newblock Initial coalescence of a drop at a planar liquid surface.
\newblock Physical Review E. 2019;100(3):033112.

\bibitem{Aryafar2008hydrodynamic}
Aryafar H, Kavehpour H.
\newblock Hydrodynamic instabilities of viscous coalescing droplets.
\newblock Physical Review E. 2008;78(3):037302.

\bibitem{deMalmazet2015coalescence}
De~Malmazet E, Risso F, Masbernat O, Pauchard V.
\newblock Coalescence of contaminated water drops at an oil/water interface:
  Influence of micro-particles.
\newblock Colloids and Surfaces A: Physicochemical and Engineering Aspects.
  2015;482:514--528.

\end{thebibliography}
\end{document}